\documentclass[12pt]{JHEP3}
\input epsf.tex
\epsfclipon
\usepackage{epsfig}
\usepackage{epstopdf}
\usepackage{epsf}
\input{epsf.sty}
\usepackage{graphicx,amsmath,amssymb}
\usepackage{epsfig,multicol}
\allowdisplaybreaks

\usepackage{bbm,bm,amsmath,amssymb}

\def\bg{\begin{eqnarray}}
\def\nd{\end{eqnarray}}
\def\sin{{\rm sin}}
\def\cos{{\rm cos}}
\def\tan{{\rm tan}}

\title{Non-K\"ahler Resolved Conifold, Localized Fluxes in M-Theory and Supersymmetry}
\author{Keshav Dasgupta${}^{1, 2}$, Maxim Emelin${}^{1}$ and Evan McDonough${}^{1}$\\
\vskip.03in
${}^1$Ernest Rutherford Physics Building, McGill University,\\
~~3600 University Street, Montr{\'e}al QC, Canada H3A 2T8\\
${}^2$Varian Physics Building, Stanford University,\\
~~382 Via Pueblo Mall, Stanford CA, USA 94305-4060\\ 
~~{\tt keshav@hep.physics.mcgill.ca, maxim.emelin@mail.mcgill.ca}\\
~~{\tt evanmc@mcgill.ca}}

\date{\today}
\maketitle

\abstract{The known supergravity solution for wrapped D5-branes on the two-cycle of a K\"ahler resolved conifold is in general ISD but not 
supersymmetric, with the supersymmetry being broken by the presence of (1, 2) fluxes. However if we allow a non-K\"ahler metric on the resolved conifold, 
supersymmetry can easily be restored. The vanishing of the (1, 2) fluxes here corresponds to, under certain conformal rescalings of the metric,
the torsion class constraints. We construct a class of explicit non-K\"ahler metrics on the resolved conifold satisfying the constraints.
All this can also be studied from M-theory, where the fluxes and branes become non-localized G-fluxes on deformed Taub-NUT spaces.
Interestingly, the gauge fluctuations
on the wrapped D5-branes appear now as localized G-fluxes in M-theory. These localized fluxes are related to certain harmonic two-forms that are normalizable. We compute 
these forms explicitly and discuss how new constraints on the geometry of the non-K\"ahler manifolds may appear from M-theory considerations.}
\begin{document}

\section{Introduction}

The resolved conifold, originally discussed in the work of \cite{candelas} supports, for a given complex structure and a given K\"ahler class, a unique Ricci flat metric with 
vanishing first Chern class. This is the well known Calabi-Yau metric that has been used in string theory to understand various aspects of dualities and compactifications.
However, in most of the studies the effect of the background fluxes on the metric of the K\"ahler resolved conifold has not been discussed. In certain interesting cases, which will be the subject of this paper, the combined effect of string equations of motion and supersymmetry may lead to metrics on the resolved conifold that are neither  K\"ahler nor Calabi-Yau. These non-K\"ahler metrics on a resolved conifold have not been used much in the literature, despite their apparent ubiquity, partly because they 
do not satisfy the nice properties one encounters for the K\"ahler case, and partly because of their underlining technicalities.

The situation changed once simple examples of non-K\"ahler manifolds that satisfy torsional equations and supersymmetry \cite{strom} were constructed in \cite{DRS}. This was followed 
by the classic work of Chiossi and Salamon \cite{chiossi} who essentially laid out the criteria for constructing torsional solutions. The condition for 
supersymmetry of these solutions were replaced, from the standard $SU(3)$ and $G_2$ holonomies \cite{CHSW}, to the corresponding $SU(3)$ and $G_2$ structures. In simpler terms, 
the non-closure of the fundamental two form or the holomorpic three-form were measured by the torsion classes: basically telling us that there are five torsion classes 
${\cal W}_i$ that form 
various representations of an $SU(3)$ structure. The well-know Calabi-Yau case arises when all the torsion classes vanish.
In M-theory, the equivalent picture with $G_2$ structure leads to four torsion classes. 

The work of Chiossi and Salamon were immediately applied to string theory in a series of beautiful papers \cite{Lopes Cardoso:2002hd, louis, lust2, gauntlett} that clarified the 
role of complexity, torsion, supersymmetry and their interconnections in constructing six dimensional manifolds in string theory. One of the early outcomes of 
this was the realization that the internal six-dimensional manifolds do not have to be K\"ahler or even complex to satisfy stringy EOM or supersymmetry. If the manifolds
are endowed with an almost complex structure that is non-integrable, consistent solutions can be constructed. The earlier work of \cite{DRS} argued the existence of 
a six-dimensional compact space that has an integrable complex structure but is non-K\"ahler. Combining the stories together, we can now construct internal spaces that are 
neither K\"ahler nor complex, yet preserve supersymmetry.  
 
One may also go back to familiar territory, for example the resolved conifold, and ask if it is possible to put a non-K\"ahler metric on it. This is the subject of this paper, and
our answer will be in the affirmative: we will be able to construct explicit non-K\"ahler metrics on a resolved conifold in sec. \ref{sec5}. In fact in sec. \ref{exsol}
we will argue that there is a large class of possible solutions with and without dilaton profiles. 

Our key result, presented in section \ref{exsol}, is two classes of explicit solutions for the supersymmetric non-K\"ahler resolved conifold, with metric given by equation \eqref{iibform}:
\bg
&&ds^2 = {1\over e^{2\phi/3} \sqrt{e^{2\phi/3} + \Delta}} ~ds^2_{0123} + e^{2\phi/3} \sqrt{e^{2\phi/3} + \Delta} ~ds^2_6 ,  \\
&& ds^2_6 = F_1~ dr^2 + F_2 (d\psi + {\rm cos}~\theta_1 d\phi_1 + {\rm cos}~\theta_2 d\phi_2)^2  + \sum_{i = 1}^2 F_{2+i}
(d\theta_i^2 + {\rm sin}^2\theta_i d\phi_i^2) . \nonumber 
\nd
We derive a class of solutions with constant dilaton, with warp factors given by:
\bg
&& F_2 = F_2(r), \;\;\;\;\;\; \phi=\phi_0 , \\
&& F_1 =\frac{4 F_{2r}^2}{F_2^{5/3} (3 + 2F_2)^{10/3}}, ~~~F_3 = 1 -\frac{2+ F_2}{F_2^{1/3} (3 + 2F_2)^{2/3}} , \nonumber\\
&&F_4= \left[1 -\frac{2 + F_2}{F_2^{1/3} (3 + 2F_2)^{2/3}}\right]{F^{2/3}_2\over (3 + 2F_2)^{2/3}} ,\nonumber \nd
where $F_{2r}= dF_2/dr$, in terms of a general $F_2(r)$, as well as a class of solutions with varying dilaton, given by:
\bg\label{churamont}
F_1 = {1\over 2F}, ~~~~~ F_2 = {r^2 F\over 2}, ~~~~~F_3 = {r^2\over 4} + a^2 e^{-2\phi}, ~~~~~ F_4 = {r^2\over 4}, ~~~~~ \phi=\phi(r) , \nd
for a general function $F(r)$. In this class we also discuss briefly the existence of 
another related background whose warp factors differ slightly from 
\eqref{churamont} by the dilaton factor:
\bg\label{suramon} 
F_1 = {e^{-\phi}\over 2F}, ~~~~~ F_2 = {r^2 e^{-\phi} F\over 2}, ~~~~~F_3 = {r^2 e^{-\phi}\over 4} + {\cal O}(a^2), ~~~~~ 
F_4 = {r^2 e^{-\phi}\over 4}. \nd
There are many advantages in constructing non-K\"ahler metrics on a resolved conifold, and in the following we mention a few. The foremost is the study of gauge/gravity duality 
in the geometic transition setting \cite{vafa}, where the starting point is the gauge theory on wrapped D5-branes on the two-cycle of a resolved conifold. As discussed in 
\cite{bec1, bec2, bec3, fangpaul}, the metric on the resolved conifold has to be non-K\"ahler to satisfy all the EOMs and supersymmetry. In general wrapping D5-branes 
on a Calabi-Yau resolved conifold would lead to a background that satisfies EOM but breaks supersymmetry \cite{pandoz, cvetic, anke, sully}. Using the criteria of \cite{GKP}, 
this means the 
background (admitting an integrable complex structure), will have a (1, 2) piece in addition to the supersymmetry preserving (2, 1) piece of the three-form flux (see 
\cite{cvetic, anke, sully} for more details). 

A related construction with non-K\"ahler resolved conifold appears when we take the mirror of the wrapped D5-branes background. The mirror, or SYZ transformation \cite{syz}, leads 
to a non-K\"ahler deformed conifold with D6-branes wrapping the three-cycle of the manifold. Under geometric transition \cite{vafa} this will give us another non-K\"ahler resolved 
conifold with fluxes in type IIA theory. The difference now is, other than the fact that we are in type IIA instead of type IIB theory, that there are no wrapped branes. The branes have
disappeared and are replaced by fluxes \cite{vafa, bec1, fangpaul}.   

Another place where non-K\"ahler resolved conifold shows up is in the gravity dual of little string theories (LSTs) recently studied in \cite{lapan}. The LSTs are constructed 
in $SO(32)$ and $E_8 \times E_8$ heterotic theories by wrapping heterotic five-branes on the two-cycle of a non-K\"ahler resolved conifold. These LSTs are not scale invariant, and
their degrees of freedom confine at low energies. In our class of solutions, the backgrounds studied in \cite{lapan} fall in the category with non-trivial dilaton profiles. 
In fact we study two kinds of solutions with varying dilaton profiles as mentioned above and exemplified by \eqref{churamont} and \eqref{suramon}; and the ones 
studied in \cite{lapan} fall in the 
latter category. 

The non-K\"ahler resolved conifold has also appeared in the study of large $N$ thermal QCD, that confines in the far IR and becomes scale invariant at the highest energies 
\cite{mia, aalok}. The background to study QGP properties involves the resolved conifold\footnote{More precisely, the small $r$ background 
should be a warped resolved-deformed conifold, which is another non-K\"ahler manifold. The deformation parameter is related to the confinement 
scale and the resolution parameter is related to scale at which UV degrees of freedom enter the system. For small confinement scale and 
large UV scale, it is the non-K\"ahler resolved conifold that would capture the dynamics succinctly. See \cite{mia, mia2} for details.}  
with three-form fluxes in type IIB theories, but in general there are no wrapped 
D5-branes. However wrapped anti-D5 branes appear once we demand UV completions with asymptotic AdS spaces \cite{mia2}, and the full construction becomes more involved than the simple 
cases that we discuss here. Nevertheless, the starting point is still a non-K\"ahler resolved conifold. 
 
Last, but not the least, the non-K\"ahler resolved conifold can also be used to generate D-terms in type IIB theory using an embedding of D7-brane in this background \cite{sully}. The subtlety 
here is to generate supersymmetry breaking bulk fluxes that are {\it not} ISD to allow for non-zero F-terms, as ISD (1,2) fluxes will break supersymmetry without generating a 
bulk cosmological constant. This criteria is essential, otherwise no D-terms could appear in the theory \cite{nilles, mia3}. These D-terms appear from certain `localized fluxes' in 
M-theory on a four-fold. Locally the four-fold will be a Taub-NUT fibered over a four-dimensional base. 

On the other hand, we can also study localized fluxes on a {\it seven}-dimensional manifold in M-theory that is locally a Taub-NUT fibered over a three-dimensional base. We expect the 
seven-dimensional manifold to have a $G_2$ structure, and appear from a specific ${\bf S}^1$ fibration over the non-K\"ahler resolved conifold. These two descriptions should match up, with the $G_2$ structure manifold being constructed by T-dualizing the type IIB solution, and by rewriting it as a warped Taub-NUT fibration over a three-dimensional base. The localized fluxes then are related to certain harmonic two-form on the warped Taub-NUT space.   
      
The paper is organized in the following way. In sec. \ref{sec2} we study the basic construction of fluxes and D5-branes on a non-K\"ahler resolved conifold using various dualities,
and then in sec. \ref{susic} argue for supersymmetry and corresponding constraints on the warp factors. The issue of supersymmetry is dealt with again in sec. \ref{sec3}, now using
the detailed machinery of the torsion classes both before, in sec. \ref{befdual}, and after, in sec. \ref{aftdual}, certain solution-generating duality transformations. The system
is then lifted to M-theory in sec. \ref{sec4} where Taub-NUT spaces appear prominently. Simple warm-up examples for generating localized fluxes using harmonic two-forms 
are discussed in sec. \ref{warm1} and sec. \ref{warm2}. We
head on to the explicit construction of solutions in sec. \ref{sec5}, and in sec. \ref{exsol} solutions for the warp factors of non-K\"ahler resolved conifolds for the two class of examples, 
with and without dilaton profiles, are derived. In sec. \ref{disc} we discuss localized fluxes and DBI gauge fields on the brane worldvolume, and conclude in sec. \ref{concl} with some discussion on the directions for future work.

\newpage

\section{D5-branes on a Non-K\"ahler Resolved Conifold \label{sec2}}

The supersymmetric case of a wrapped D5-brane on a resolved conifold is not hard to construct if we assume that the metric 
on a resolved conifold is a non-K\"ahler one. This has been discussed in a different context in \cite{MM, fangpaul}, and we 
will first outline the general technique. The starting point is a non-K\"ahler resolved cone solution in type IIB in the 
presence of $H_{NS}$ fluxes. This is supersymmetric and the solution is given by the following form:
\bg\label{kunku}
&& ds^2 = ds^2_{0123} + e^{2\phi} ds_6^2 , \nonumber\\
&& {\cal H} = e^{2\phi} \ast_6 d\left(e^{-2\phi} J\right) ,\nd
where $\phi$ is the usual type IIB dilaton and $J$ is the fundamental two-form of the 
warped internal six-dimensional manifold whose unwarped metric
is given by:
\bg\label{nonkah}
ds^2_6 = F_1~ dr^2 + F_2 (d\psi + {\rm cos}~\theta_1 d\phi_1 + {\rm cos}~\theta_2 d\phi_2)^2  + \sum_{i = 1}^2 F_{2+i}
(d\theta_i^2 + {\rm sin}^2\theta_i d\phi_i^2) ,\nonumber\\
\nd
where $F_i$ are warp-factors whose values will be determined later. For simplicity we will consider them to be functions of the radial coordinate $r$ only.

The steps for creating a supersymmetric wrapped D5 brane with three-form fluxes now follow the trick laid out by \cite{MM, fangpaul}. 
We S-dualize the background \eqref{kunku}, followed by three T-dualities along the $x^{1, 2, 3}$ directions. The resulting
type IIA configuration will now become:
\bg\label{kunkumeno}
&&ds^2 = - e^{-\phi} dt^2 + e^{\phi} ds^2_{123} + e^{\phi} ds^2_6 ,\nonumber\\
&& G_4 = d(e^{-2\phi} J) \wedge dt , \nd
with a dilaton expressed as $e^{\phi/2}$. We can lift the configuration \eqref{kunkumeno} to M-theory and perform a boost along the
eleventh direction. Using the boost parameter $\beta$, the resulting M-theory configuration is given by:
\bg\label{jhotka}
&&ds^2 = -dt^2 (e^{-4\phi/3} - \Delta)  + dx_{11}^2 (e^{2\phi/3} + \Delta) + e^{2\phi/3}\left(ds^2_{123} + ds^2_6\right) , \nonumber\\
&& G_4 = \left(G_4\right)_{0mnp} {\rm cosh}~\beta ~ dt \wedge dx^m \wedge dx^n \wedge dx^p   - 
\left(G_4\right)_{0mnp} {\rm sinh}~\beta ~ dx_{11} \wedge dx^m \wedge dx^n \wedge dx^p , \nonumber\\
\nd
where ($m, n, p$) in the subscript of $G_4$ denote the coordinates of the internal non-K\"ahler manifold. We have also defined $\Delta$ as:
\bg\label{linfri}
\Delta = {\rm sinh}^2\beta \left(e^{2\phi/3} - e^{-4\phi/3}\right) , \nd 
which vanishes when there is no dilaton or no boost as 
expected. 

Once we dimensionally reduce this to type IIA and then make the three T-dualities along directions $x^{1, 2, 3}$, the 
type IIB configuration takes the following form:
\bg\label{iibform}
&&ds^2 = {1\over e^{2\phi/3} \sqrt{e^{2\phi/3} + \Delta}} ~ds^2_{0123} + e^{2\phi/3} \sqrt{e^{2\phi/3} + \Delta} ~ds^2_6 , \nonumber\\
&& {\cal F}_3 = {\rm cosh}~\beta e^{2\phi} \ast_6 d\left(e^{-2\phi} J\right), ~~~~~~ {\cal H}_3 = -{\rm sinh}~\beta ~d\left(e^{-2\phi} J\right) , \nd
with dilaton $e^{\phi_B} = e^{-\phi}$ and the Hodge-star above is with respect to the non-K\"ahler metric \eqref{kunku}. Note that if the underlying
metric on the resolved conifold was K\"ahler, then it {\it wouldn't} have been possible to have a 
supersymmetric configuration like \eqref{kunkumeno}. The above metric describes D5 branes wrapped on the warped resolved conifold, the detailed study of which is the subject of our present work.

We will also have a five-form given by:
\bg\label{5form}
{\widetilde{\cal F}}_5 = -{\rm sinh}~\beta~{\rm cosh}~\beta\left(1 + \ast_{10}\right) {\cal C}_5(r)~d\psi \wedge \prod_{i=1}^2~\sin~\theta_i~ d\theta_i \wedge d\phi_i\nd
where the Hodge-star is now with respect to the metric \eqref{iibform} and ${\cal C}_5(r)$ will be determined below. 
Thus combining \eqref{iibform} and \eqref{5form} we should get our supersymmetric
background. Also note that the 3-form fluxes obey a {\it modified} ISD condition:
\bg\label{isd}
{\cal F}_3 + e^{2\phi} {\rm tanh}~\beta~\ast_6 {\cal H}_3 = 0 . \nd
This guarantees the configuration is a solution to the equations of motion, but does not guarantee supersymmetry. 
For supersymmetry we will need extra conditions on the warp factors.
We will discuss this in sec. \ref{susic}.

It will also
be useful to expand the ${\cal H}_3$ and ${\cal F}_3$ forms from \eqref{iibform} in terms of the coordinate one-forms. Using the metric \eqref{nonkah},
we find ${\cal H}_3$ to be given by the following expression:
\bg\label{boklik}
{{\cal H}_3\over {\rm sinh}~\beta} &= &~~\Big(\sqrt{F_1F_2}\sin\theta_1-F_{3r}\sin\theta_1\Big)dr\wedge d\theta_1\wedge d\phi_1\nonumber\\
&&+ \Big(\sqrt{F_1F_2}\sin\theta_2-F_{4r}\sin\theta_2\Big)dr\wedge d\theta_2\wedge d\phi_2 , \nd
in terms of $F_i(r)$, $F_{nr} \equiv dF_n/dr$, $\phi_r \equiv d\phi/dr$, and we have assumed $\phi=\phi(r)$ to be 
independent of the angular coordinates. The RR three-form ${\cal F}_3$ also simplifies when the dilaton is 
independent of the angular coordinates, taking the following form:
\bg\label{hingsro}
{{\cal F}_3\over {\rm cosh}~\beta} = &&~~k_1 F_2 \cos~\theta_2(\sqrt{F_1F_2}
- F_{4r})d\theta_1\wedge d\phi_1\wedge d\phi_2\nonumber\\
&&+~ k_2~F_2 \cos~\theta_1(\sqrt{F_1F_2} -F_{3r})d\theta_2\wedge d\phi_1\wedge d\phi_2\nonumber\\
&&+ ~k_3~\sin~\theta_2(\sqrt{F_1F_2} -F_{4r})d\psi\wedge d\theta_1\wedge d\phi_1\nonumber\\
&&+ ~k_4~\sin~\theta_1(\sqrt{F_1F_2}-F_{3r})d\psi\wedge d\theta_2\wedge d\phi_2 ,
\nd
where again $F_{nr} \equiv dF_n/dr$ and the $k_i$ are defined by the following expressions:
\bg\label{kidefn} 
&&k_1 = - {F_3 e^{2\phi}\over F_4 \sqrt{F_1 F_2}} \cdot\sin~\theta_1,
~~~~~k_2 =  {F_4 e^{2\phi}\over F_3 \sqrt{F_1 F_2}}\cdot\sin~\theta_2 , \nonumber\\
&&k_3 = - e^{2\phi}\sqrt{F_2\over F_1}\cdot{F_3\over F_4}
\cdot{\sin~\theta_1\over \sin~\theta_2},~~~~~
k_4 = - e^{2\phi}\sqrt{F_2\over F_1}\cdot{F_4\over F_3}\cdot{\sin~\theta_2\over \sin~\theta_1} .
\nd
Note that $d{\cal H}_3 = 0$, whereas $d{\cal F}_3$ does \emph{not} vanish: an indication that there are wrapped five-brane sources. Finally using 
\eqref{hingsro} and \eqref{boklik}, ${\cal C}_5(r)$ in \eqref{5form} can be expressed as:
\bg\label{juasha}
{\cal C}_5(r) = \int^r~{e^{2\phi} F_3 F_4 \sqrt{F_1 F_2}\over F_1}\left[\left({\sqrt{F_1F_2} - F_{3r}\over F_3}\right)^2 
+ \left({\sqrt{F_1F_2} - F_{4r}\over F_4}\right)^2\right] dr . \nd
To lift this configuration to M-theory, we will first T-dualize along $\psi$ direction to generate a six-brane configuration in 
IIA\footnote{Depending on what energy scale we are in, T-duality may lead to either the full D6-brane or a D4-brane. See the discussion around eq. \eqref{c7} where subtleties associated with the T-duality is discussed.}. 
The IIB $\mathcal{F}_3$ gives rise to the following RR two-form flux in the dual type IIA theory:
\bg\label{gaugo} 
{\cal F}_2 =&& ~- e^{2\phi}\sqrt{F_2\over F_1}\cdot{F_3\over F_4}(\sqrt{F_1F_2} -F_{4r}) {\rm cosh}~\beta~
\sin~\theta_1 d\theta_1\wedge d\phi_1\nonumber\\
 && ~- e^{2\phi}\sqrt{F_2\over F_1}\cdot{F_4\over F_3}(\sqrt{F_1F_2} -F_{3r}){\rm cosh}~\beta~\sin~\theta_2 d\theta_2\wedge d\phi_2 .\nd
We will impose non-closure of ${\cal F}_2$ to allow for a Taub-NUT background along the angular directions in the 
lift to M-theory. We also define a warp factor
$H$ in the following way from \eqref{iibform}:
\bg\label{heccu} 
H = e^{4\phi/3} \left(e^{2\phi/3} + \Delta\right) , \nd
where $\Delta$ is defined in \eqref{linfri}. Using this, the M-theory metric can be expressed as:
\bg\label{hingsro2}
ds^2_{11} &=& {e^{2\phi/3} F_2^{1/3}\over H^{1/3}}\left(ds^2_{0123} + {1\over F_2} d\psi^2\right) + {1\over e^{4\phi/3}F_2^{2/3}H^{1/3}}
(dx_{11} + {\cal A}_{1\mu} dx^\mu)^2 \nonumber\\
&+& e^{2\phi/3}F_2^{1/3} H^{2/3}\left[F_1 dr^2 + F_3(d\theta_1^2 + \sin^2\theta_1~d\phi_1^2) + 
F_4(d\theta_2^2 + \sin^2\theta_2~d\phi_2^2)\right]\nonumber\\
&=& G_1\left(ds^2_{0123} + {1\over F_2}d\psi^2\right) + G_2\left(d\theta_1^2 + {\rm sin}^2\theta_1~d\phi_1^2\right)\nonumber\\
&+& G_3 dr^2 + G_4 \left(d\theta_2^2 + {G_5\over G_4} d\phi_2^2\right) + G_6\left(dx_{11} + {\cal A}_{1\mu} dx^\mu\right)^2 , \nd
where ${\cal A}_1$ is the gauge-field coming from \eqref{gaugo}.
This is a deformed Taub-NUT geometry that stretches along directions 
($r, \theta_2, \phi_2, x_{11}$). We will discuss this geometry in more details in sec. \ref{sec4}.
Additionally, comparing \eqref{hingsro2} to \eqref{juta} we see that the relevant $G_i$ coefficients are 
defined as:
\bg\label{lfiday}
&& G_1 = e^{2\phi/3} F_2^{1/3} H^{-1/3}, ~~~~~~~~~~~~~~~~G_2= e^{2\phi/3} F_2^{1/3} H^{2/3} F_3 ,\nonumber\\
&& G_3 = e^{2\phi/3} H^{2/3} F_1 F_2^{1/3}, ~~~~~~~~~~~~~~ G_4 = e^{2\phi/3} H^{2/3} F_4 F_2^{1/3} ,\nonumber\\ 
&& G_5 = e^{2\phi/3} H^{2/3} F_4 F_2^{1/3} \sin^2\theta_2, ~~~~~~~ G_6 = {e^{-4\phi/3} H^{-1/3} F_2^{-2/3}}.\nd
There is also background $G$-flux ${\cal G}_4$ to support this configuration. This is given by:
\bg\label{lrus}
{{\cal G}_4 \over {\rm sinh}~\beta} = && ~~\left(\sqrt{F_1 F_2} - F_{3r}\right) \sin~\theta_1 ~dr\wedge d\theta_1 \wedge d\phi_1 \wedge dx_{11} +
{\sin~\theta_1\over {\rm sinh}~\beta} ~d\psi \wedge d\theta_1 \wedge d\phi_1 \wedge dx_{11}\nonumber\\
&& + \left(\sqrt{F_1 F_2} - F_{4r}\right) \sin~\theta_2 ~dr\wedge d\theta_2 \wedge d\phi_2 \wedge dx_{11}
+ {\sin~\theta_2\over {\rm sinh}~\beta} ~d\psi \wedge d\theta_2 \wedge d\phi_2 \wedge dx_{11} \nonumber\\
&& +~ k_1 F_2 \left(\sqrt{F_1 F_2} - F_{4r} \right) {\rm coth}~\beta~
\cos~\theta_2 ~d\theta_1 \wedge d\phi_1 \wedge d\phi_2 \wedge d\psi\nonumber\\
&& +~ k_2 F_2 \left(\sqrt{F_1 F_2} - F_{3r} \right) {\rm coth}~\beta~
\cos~\theta_1 ~d\theta_2 \wedge d\phi_1 \wedge d\phi_2 \wedge d\psi\nonumber\\
&& +~ {\cal C}_1(r, \theta_1, \theta_2)~dr \wedge d\theta_1 \wedge d\phi_1\wedge d\phi_2 
+ {\cal C}_2(r, \theta_1, \theta_2)~dr \wedge d\theta_2 \wedge d\phi_2\wedge d\phi_1,\nd
where the last two terms are from the five-form \eqref{5form} with the coefficients ${\cal C}_i$ derivable directly from \eqref{5form} and 
\eqref{juasha}. We will discuss 
more on this soon.
 
\subsection{Supersymmetry Constraints on the Warp Factors \label{susic}}

We can now explicitly demonstrate supersymmetry for the type IIB background with D5-branes wrapping a resolved conifold, equation \eqref{iibform}. This requires the ${\cal G}_3$ flux:
\bg\label{rehed}
{\cal G}_3 = {\cal F}_3 - i e^{-\phi_B} {\cal H}_3 , \nd
to be of a (2, 1) form, and not a (1, 2) form. We will first need the vielbeins for the background \eqref{iibform}. They are given by:
\bg\label{vielbeins}
&& e_1 = \sqrt{F_1\sqrt{H}}e_r, ~~~~~~ e_2 = \sqrt{F_2\sqrt{H}}(d\psi + \cos~\theta_1 d\phi_1 + \cos~\theta_2 d\phi_2) = \sqrt{F_2\sqrt{H}} e_\psi , \nonumber\\
&& e_3 = \sqrt{F_3\sqrt{H}}\left(-\sin~{\psi\over 2} ~e_{\phi_1} + \cos~{\psi\over 2}~e_{\theta_1}\right), 
~~e_4 = \sqrt{F_3\sqrt{H}}\left(\cos~{\psi\over 2} ~e_{\phi_1} + \sin~{\psi\over 2}~e_{\theta_1}\right) , \nonumber\\   
&& e_5 = \sqrt{F_4\sqrt{H}}\left(-\sin~{\psi\over 2} ~e_{\phi_2} + \cos~{\psi\over 2}~e_{\theta_2}\right), 
~~e_6 = \sqrt{F_4\sqrt{H}}\left(\cos~{\psi\over 2} ~e_{\phi_2} + \sin~{\psi\over 2}~e_{\theta_2}\right) , \nonumber\\ \nd
where $H$ is the warp factor \eqref{heccu}. Now using the vielbeins we can define three complex one-forms in the following way:
\bg\label{c1fom}
E_1 = e_1 + i\gamma e_2, ~~~~~~~ E_2 = e_3 + i e_4, ~~~~~~~ E_3 = e_5 + i e_6 ,\nd
where we have inserted a non-trivial complex structure ($i\gamma, i, i$) respectively. The functional form of $\gamma$ will be derived soon.
Using the complex one-forms \eqref{c1fom}, we can rewrite the ${\cal G}_3$ flux \eqref{rehed} in the following way:
\bg\label{choddal} 
{\cal G}_ 3&=& -{1\over 4}\left[{e^{\phi}(\sqrt{F_1F_2} - F_{3r}){\rm sinh}~\beta \over F_3 \sqrt{H}\sqrt{F_1\sqrt{H}}} 
- {e^{2\phi}(\sqrt{F_1F_2} -F_{4r}) {\rm cosh}~\beta \over
\gamma F_4\sqrt{H} \sqrt{F_1\sqrt{H}}}\right] ~E_1 \wedge E_2 \wedge \bar{E}_2\nonumber\\
&& +{1\over 4}\left[{e^{\phi}(\sqrt{F_1F_2} - F_{3r}){\rm sinh}~\beta \over F_3\sqrt{H} \sqrt{F_1\sqrt{H}}} 
+ {e^{2\phi}(\sqrt{F_1F_2} -F_{4r}) {\rm cosh}~\beta\over
\gamma F_4\sqrt{H} \sqrt{F_1\sqrt{H}}}\right] ~E_2 \wedge {\bar E}_1 \wedge \bar{E}_2\nonumber\\
&&- {1\over 4}\left[{e^{\phi}(\sqrt{F_1F_2} - F_{4r}){\rm sinh}~\beta \over F_4 \sqrt{H} \sqrt{F_1\sqrt{H}}} 
- {e^{2\phi}(\sqrt{F_1F_2} -F_{3r}) {\rm cosh}~\beta \over
\gamma F_3 \sqrt{H} \sqrt{F_1\sqrt{H}}}\right] ~E_1 \wedge E_3 \wedge \bar{E}_3\nonumber\\
&&+ {1\over 4}\left[{e^{\phi}(\sqrt{F_1F_2} - F_{4r}){\rm sinh}~\beta \over F_4 \sqrt{H}\sqrt{F_1\sqrt{H}}} 
+ {e^{2\phi}(\sqrt{F_1F_2}-F_{3r}){\rm cosh}~\beta  \over
\gamma F_3 \sqrt{H} \sqrt{F_1\sqrt{H}}}\right] ~E_3 \wedge {\bar E}_1 \wedge \bar{E}_3. \nonumber\\
\nd
We note that the above rewriting of the ${\cal G}_3$ flux shows that there are both ($2, 1$) as well as ($1, 2$) pieces, although the total 
flux is ISD. However we can make the ($1, 2$) piece vanish by choosing an appropriate $\gamma$. 

A careful look at \eqref{choddal} tells us that the 
vanishing of the $E_2 \wedge {\bar E}_1 \wedge {\bar E}_2$ term requires $\gamma$ to be:
\bg\label{gamma212}
\gamma = -{e^{\phi}\left(\sqrt{F_1F_2} - F_{4r}\right)\over \sqrt{F_1F_2} - F_{3r}}\cdot {F_3\over F_4} ~{\rm coth}~\beta ,\nd
whereas vanishing of the $E_3 \wedge {\bar E}_1 \wedge {\bar E}_3$ part requires $\gamma$ to be:
\bg\label{gamma313}
\gamma = -{e^{\phi}\left(\sqrt{F_1F_2} - F_{3r}\right)\over \sqrt{F_1F_2} - F_{4r}}\cdot {F_4\over F_3} ~{\rm coth}~\beta . \nd
It is easy to see that both \eqref{gamma212} and \eqref{gamma313} are satisfied when we choose $\gamma$ to be the following:
\bg\label{equal}
\gamma = \pm e^\phi ~{\rm coth}~\beta , \nd
which allows us to choose the complex structure as ($\pm ie^\phi ~{\rm coth}~\beta, i, i$) for our choice of vielbeins.

Note that this analysis is only valid if the internal manifold is complex. In our case, this requires a constant dilaton, $\phi=\phi_0$, which can be seen by computing the Nijenhuis tensor, and also from the torsion class analysis that we will present in sec. \ref{sec3}. Of course, if we relax this condition on the dilaton we can allow for non-complex manifolds, however we will not do just yet, and will instead take $\gamma$ to be:
\bg\label{kkbacca}  
\gamma = \pm e^{\phi_0}~{\rm coth}~\beta . \nd
Let us first consider the option with a minus sign in \eqref{equal}.  This choice leads us to the following constraint
on the warp factors $F_3$ and $F_4$:
\bg\label{juddho}
{\sqrt{F_1 F_2} - F_{3r}\over \sqrt{F_1 F_2} - F_{4r}} =  {F_3\over F_4} . \nd
A sub-class of solutions satisfying \eqref{juddho}, for the choice of 
\eqref{equal}, will be the case where the warp factors satisfy: 
\bg\label{lattu}
F_3(r) ~ = ~ F_4(r) \nd
which corresponds to a {\it singular} non-K\"ahler conifold geometry, provided ($F_3, F_4$) vanish at $r = 0$. In general this equality cannot hold for the 
resolved conifold case as we expect locally $F_3 - F_4 = a^2$ where $a^2$ is the constant resolution parameter. To allow for a resolved conifold one could in principle demand:
\bg\label{jisha}
F_{3r} ~ = ~ F_{4r} ,\nd
however such a choice leads to a contradiction, unless we put $F_3 = F_4$ and ($F_3, F_4$) now non-vanishing at $r = 0$. 
The underlying reason for this is because our case is restrictive, i.e. we have made all the warp factors functions of the radial coordinate $r$ only. If we keep the warp factors functions of the angular coordinates ($\theta_1, \theta_2$), we will not have to impose the equality \eqref{lattu}. However we will not do the most generic case
here. 

For the moment, let's proceed with the singular conifold background \eqref{lattu}, and compute the RR gauge field \eqref{gaugo}. Under the constraint \eqref{lattu} the form of the gauge field changes from
\eqref{gaugo} to the following:
\bg\label{gaugonow}
{\cal F}_2 = - \sqrt{F_2\over F_1} \cdot \left(\sqrt{F_1 F_2} - F_{4r}\right) {\rm cosh}~\beta 
\left(e_{\theta_1} \wedge e_{\phi_1} + e_{\theta_2} \wedge e_{\phi_2}\right) .  \nd
One simple solution for the Taub-NUT to 
be along the angular directions, i.e. for closed ${\cal F}_2$, is that the warp factors $F_i$ satisfy the following differential equation:
\bg\label{wadidiff}
{dF_4\over dr} = \sqrt{F_1 F_2}\left(1- {e^{-2\phi_0} \over {F_2}}\right) . \nd
The above relation, together with \eqref{lattu} and \eqref{jisha}, succinctly summarizes the constraints on the warp factors of the internal manifold 
\eqref{nonkah} and the dilaton $\phi$ to allow for supersymmetric solutions of the form \eqref{iibform} with a non-K\"ahler singular conifold.

The RR gauge field for the singular conifold, using the conditions \eqref{wadidiff}, takes the form:
\bg\label{jome}
{\cal A}_1 = {\rm cosh}~\beta \left(\cos~\theta_1~d\phi_1 + \cos~\theta_2~d\phi_2\right) . \nd
At any given point on the base manifold parametrized by ($\theta_1, \phi_1, \psi$) and $x^{0, 1, 2, 3}$ the gauge field is:
\bg\label{leanunu}
{\cal A}_1 = {\rm cosh}~\beta ~\cos~\theta_2~d\phi_2 , \nd 
which is the familiar Taub-NUT form as expected. Thus our Taub-NUT space could be thought of as fibered over the two-dimensional sphere ($\theta_1, \phi_1$).

To study the \emph{resolved conifold}, let's consider a second  choice for $\gamma$, which is the plus sign solution in \eqref{equal}. For this case the constraint equation will change from
\eqref{juddho} to:
\bg\label{ghonti}
{\sqrt{F_1 F_2} - F_{3r}\over F_3} +  {\sqrt{F_1 F_2} - F_{4r} \over F_4} = 0 . \nd
Now of course \eqref{lattu} or \eqref{jisha} cannot be implemented\footnote{The case $F_3 = - F_4$ would lead to an inconsistency in \eqref{ghonti}. This makes sense as the warp factors
cannot be negative.}. However
we can still impose a slight variant of \eqref{wadidiff}, but now for $F_3$ as:
\bg\label{whatif}
{dF_3\over dr} = \sqrt{F_1 F_2}\left(1+ {e^{-2\phi_0} \over {F_2}}\right). \nd
However since $F_3 \ne \pm F_4$, we will assume:
\bg\label{tnone}
F_4 = -g_1 F_3 = \vert g_1 \vert F_3, \nd
where $g_1(r) = - \vert g_1 \vert$ is a function of $r$ satisfying the following relation in terms of the warp factors:
\bg\label{lerae}
{dg_1\over dr} = -{\sqrt{F_1F_2}\over F_3}\left[1 + g_1\left(1+ {2e^{-2\phi_0} \over F_2}\right)\right]. \nd
This is a generalization of the resolved conifold background, where the resolution parameter $(F_3 - F_4)$ is given by the function $(1 - \vert g_1 (r)\vert) F_3$. 

Given the functional forms of ($F_1, F_2, \phi$), we can compute $F_3$ from \eqref{wadidiff}. Once $F_3$ is determined, $g_1(r)$ can be found from \eqref{lerae} above, and 
knowing $g_1(r)$ will give us $F_4$ from \eqref{tnone}\footnote{Tighter constraints on the warp factors will be discussed later.}. 
Therefore after the dust settles, the background RR gauge field will change from \eqref{jome} to take the
following field strength: 
\bg\label{jomadar}
{\cal F}_2 = -{\rm cosh}~\beta \left[e_{\theta_1}\wedge e_{\phi_1}  + g_1(r) e_{\theta_2}\wedge e_{\phi_2}\right] \equiv d{\cal A}_1 + {\rm sources}. \nd
As before, at any given point on the two-sphere parametrized by ($\theta_1, \phi_1$), the gauge field \eqref{jomadar} will resemble somewhat \eqref{leanunu}
but with the following source equation:
\bg\label{sourceq}
d{\cal F}_2 = {\rm cosh}~\beta {\sqrt{F_1F_2}\over F_3}\left[1 + g_1\left(1+ {2 e^{-2\phi_0}\over F_2}\right)\right] e_r \wedge e_{\theta_2} \wedge e_{\phi_2} ,
\nd
which implies that there are delocalized sources along these directions. For more details on the delocalized sources, one may refer 
to \cite{deloc}.

\section{Torsion Classes, Complexity, and  Supersymmetry \label{sec3}}

Having discussed in details a special case of supersymmetry and other constraints in the previous section, let us analyze a more generic 
case using torsion classes for the background \eqref{kunku} and \eqref{nonkah}. We will then specialize our construction to the type IIB background 
\eqref{iibform} and argue for the consistency. To make the picture more succinct, we will divide our analysis for the type IIB background in two parts: before
duality and after duality. 

\subsection{Type IIB background before duality \label{befdual}}

The story before duality transformations begins from the background \eqref{kunku} and \eqref{nonkah}. With generic choices of the warp factors $F_i$ and 
the dilaton $e^\phi$ the manifold \eqref{nonkah} will be a non-K\"ahler manifold with an almost-complex structure that may or may not be integrable. In 
the language of torsion classes ${\cal W}_i$ \cite{chiossi} we have two key defining equations:
\bg\label{ghora}
&& dJ = {3\over 4} i \left({\cal W}_1 \overline{\Omega} - \overline{\cal{W}}_1 \Omega\right) + {\cal W}_3 + J \wedge {\cal W}_4 ,\nonumber\\
&& d\Omega = {\cal W}_1 J \wedge J + J \wedge {\cal W}_2 + \Omega \wedge {\rm Re}~{\cal W}_5 , \nd
with the following additional constraints:
\bg\label{churamon}
J \wedge {\cal W}_3 = J \wedge J \wedge {\cal W}_2 = \Omega \wedge {\cal W}_3 = 0 . \nd
Using the above constraints \eqref{churamon} and an additional condition $J \wedge \Omega = 0$, that will be consistent with our choice of ($J, \Omega$), 
it is easy to infer from \eqref{ghora} that (equation 2.8 of \cite{Lopes Cardoso:2002hd}):
\bg\label{fasa}
{\cal W}_1 J \wedge J \wedge J = d\Omega \wedge J  = J \wedge d\Omega , \nd
which will help us to determine ${\cal W}_1$ once we know the fundamental form $J$ and the (3, 0) form $\Omega$. Furthermore, from the (2, 2) part of 
$d\Omega$ we can determine ${\cal W}_2$ via:
\bg\label{bbbob}
[d\Omega]^{(2, 2)} = {\cal W}_1 J \wedge J + {\cal W}_2 \wedge J . \nd
Alternatively, all components of the Nijenhuis tensor are completely determined by the torsion classes ${\cal W}_1$ and ${\cal W}_2$, i.e
\bg\label{nunur}
{\cal W}_1 ~ \oplus ~ {\cal W}_2 .\nd
To proceed towards an explicit determination of the torsion components, we will need the complex vielbeins for the background \eqref{kunku} and \eqref{nonkah}. Using a slight variant of \eqref{vielbeins}, the complex
vielbeins now are:
\bg\label{coviel}
&&{\cal E}_1 = e^{\phi}\left(\sqrt{F_1} e_r + i \sqrt{F_2} e_\psi\right)  , \nonumber\\
&&{\cal E}_2 = e^{\phi+i\psi/2}\sqrt{F_3}\left(e_{\theta_1} + i e_{\phi_1}\right) , \nonumber\\ 
&& {\cal E}_3 = e^{\phi+i\psi/2}\sqrt{F_4}\left(e_{\theta_2} + i e_{\phi_2}\right) .
\nd
Note that these vielbeins differ from the ones discussed in \cite{pandoz, cvetic}, and we will argue that the choice \eqref{coviel} give consistent results 
for the corresponding Calabi-Yau case. Using the convention we have been using, the fundamental form (1, 1) $J$ is defined as:
\bg\label{jdef}
J &= &{\bar {\cal E}}_1 \wedge {\cal E}_1 + {\cal E}_2 \wedge {\bar {\cal E}}_2 + {\cal E}_3 \wedge {\bar {\cal E}}_3\nonumber\\
&=& 2i e^{2\phi}\left(\sqrt{F_1 F_2} ~e_r \wedge e_\psi + F_3 e_{\phi_1}\wedge e_{\theta_1} + F_4 e_{\phi_2}\wedge e_{\theta_2}\right) , \nd
such that $dJ$ will become:
\bg\label{dj}
dJ &=& 2i e^{2\phi}F_3 \left({\sqrt{F_1F_2}-F_{3r}\over F_3} - 2\phi_r\right) e_r \wedge e_{\theta_1} \wedge e_{\phi_1}\nonumber\\
&+& 2i e^{2\phi}F_4 \left({\sqrt{F_1F_2}-F_{4r}\over F_4} - 2\phi_r\right) e_r \wedge e_{\theta_2} \wedge e_{\phi_2} , \nd
implying that $d J=0$ when the following two conditions are met:
\bg\label{kahrest}
{\sqrt{F_1F_2}-F_{3r}\over F_3} = 2\phi_r, ~~~~~~ {\sqrt{F_1F_2}-F_{4r}\over F_4} = 2\phi_r . \nd
One may check that for the Calabi-Yau resolved conifold, where the warp factors $F_i$ take the following values \cite{pandoz}:
\bg\label{cymet} F_1 = {r^2 + 6a^2 \over r^2 + 9a^2}, ~~~~~F_2 = \left({r^2 + 9a^2 \over r^2 + 6 a^2}\right){r^2\over 9}, ~~~~~ F_3 = {r^2\over 6} + a^2, 
~~~~~F_4 = {r^2\over 6} , \nd
with $a$ being the resolution parameter,
\eqref{kahrest} is satisfied with a constant dilaton for both $a = 0$, the singular conifold case, and for $a \ne 0$, 
the standard resolved conifold case. 

Note that due to our choice of the complex structure ($i, i, i$), $dJ$ has only (2, 1) and (1, 2) pieces. Thus using the same 
vielbeins we can also to compute the (3, 0) form $\Omega$. For our case this is defined as:
\bg\label{omegadef}
\Omega & = & {\cal E}_1 \wedge {\cal E}_2 \wedge {\cal E}_3 \nonumber\\
&=& e^{i\psi} {\cal A}(r) e_r \wedge (e_{\theta_1} \wedge e_{\theta_2} - e_{\phi_1} \wedge e_{\phi_2} + i e_{\theta_1} \wedge e_{\phi_2} + i e_{\phi_1} \wedge e_{\theta_2})\nonumber\\
&+& ie^{i\psi} {\cal B}(r) d\psi \wedge (e_{\theta_1} \wedge e_{\theta_2} - e_{\phi_1} \wedge e_{\phi_2} + i e_{\theta_1} \wedge e_{\phi_2} + i e_{\phi_1} \wedge e_{\theta_2})\nonumber\\
&+& ie^{i\psi} {\cal B}(r) \left({\rm cot}~\theta_1 e_{\phi_1} \wedge e_{\theta_1} \wedge e_{\theta_2} + i {\rm cot}~\theta_1 e_{\phi_1} \wedge e_{\theta_1} \wedge e_{\phi_2}\right) \nonumber\\
&+& ie^{i\psi} {\cal B}(r) \left({\rm cot}~\theta_2 e_{\phi_2} \wedge e_{\theta_1} \wedge e_{\theta_2} + i {\rm cot}~\theta_2 e_{\phi_2} \wedge e_{\phi_1} 
\wedge e_{\theta_2}\right),\nd 
where ${\cal A}(r)$ and ${\cal B}(r)$ are defined in the following way:
\bg\label{calab}
{\cal A}(r) \equiv \sqrt{F_1 F_3 F_4}e^{3\phi}, ~~~~~~~{\cal B}(r) \equiv \sqrt{F_2 F_3 F_4}e^{3\phi} . \nd
Using this, one can easily show that $\Omega \wedge dJ=0$ and hence the first torsion class vanishes. In addition to this, since $\Omega$ is a $(3,0)$ form there will be no $(2,2)$ piece to $d\Omega$, and hence $\mathcal{W}_2$ vanishes as well. Hence we have:
\bg\label{compol}
\mathcal{W}_1 ~ = ~ {\cal W}_2 ~ = ~ 0.
\nd
This is as one would expect, since the warp factors are just functions of $r$ \cite{lapan}. Vanishing of these torsion classes mean that  the underlying manifold \eqref{nonkah} can allow integrable complex structures. Note that this result is independent of $\phi(r)$, and hence the the background \eqref{kunku} and \eqref{nonkah} will be complex for both a constant dilaton and  a varying dilaton.


To see how the warp factors in the metric \eqref{nonkah}
are constrained we need to investigate the other three torsion classes. The second equation in \eqref{ghora} can now be written as:
\bg\label{ghori}
d\Omega = \Omega \wedge {\rm Re}~ {\cal W}_5 . \nd
Under a chain of identifications ${\rm Re}~{\cal W}_5$ is now related to the dilaton profile in the following way \cite{strom, Lopes Cardoso:2002hd, louis, lust2, gauntlett}:
\bg\label{chitkar}
{\rm Re}~{\cal W}_5 = {1\over 8} \left(\Omega + \overline{\Omega}\right) \lrcorner \left(d\Omega + d\overline{\Omega}\right) = d {\rm log}~\vert\Omega\vert 
= -2 d\phi . \nd
Plugging \eqref{chitkar} in \eqref{ghori} we get:
\bg\label{photka} d\left(e^{-2\phi} \Omega\right) ~ = ~ 0 , \nd
which is a familiar condition for the manifold \eqref{nonkah} to have a $SU(3)$ structure. It is comforting to see that it appears here naturally.

The only remaining detail is to compute $d\Omega$ explicitly and compare the result with \eqref{photka}. This will help us to determine ${\rm Re}~{\cal W}_5$.
Using \eqref{omegadef}, we can easily compute $d\Omega$. This is given by:
\bg\label{domega}
d\Omega  = && ie^{i\psi}\left[{\cal A}(r) - {\cal B}'(r)\right]d\psi \wedge e_r \wedge (e_{\theta_1} \wedge e_{\theta_2} - e_{\phi_1} \wedge e_{\phi_2} + i e_{\theta_1} \wedge e_{\phi_2} + i e_{\phi_1} \wedge e_{\theta_2})\nonumber\\
&&~~~~~~ +e^{i\psi}\left[{\cal A}(r) - {\cal B}'(r)\right]\left({\rm cot}~\theta_1~e_{\theta_1} + {\rm cot}~\theta_2~e_{\theta_2}\right) \wedge e_r\wedge e_{\phi_1}\wedge e_{\phi_2}\nonumber\\
&&~~~~~~+ i e^{i\psi}\left[{\cal A}(r) - {\cal B}'(r)\right]\left({\rm cot}~\theta_1~e_{\phi_1} + {\rm cot}~\theta_2~e_{\phi_2}\right) \wedge e_r\wedge e_{\theta_1}\wedge e_{\theta_2} , \nd
where prime denotes derivative with respect to the radial coordinate $r$. It is now interesting to note that when the dilaton is a constant, \eqref{photka} 
$SU(3)$ structure requires $d\Omega$ to vanish. In general, the condition $d \Omega=0$ requires:
\bg\label{condu}
{\cal A}(r) = {\cal B}'(r) , \nd
which in the language of the warp factors $F_i$ and the dilaton $\phi$ gives the following constraint:
\bg\label{bgcos}
{F_{3r}\over F_3} + {F_{4r}\over F_4} + {F_{2r} - 2 \sqrt{F_1 F_2}\over F_2} + 6\phi_r ~ = ~ 0 . \nd
Again it is easy to see that the Calabi-Yau resolved conifold or the singular conifold with warp factors given in \eqref{cymet}, for $a \ne 0$ and $a = 0$ 
respectively, satisfy the constraint \eqref{condu}. Thus they are K\"ahler manifolds as expected\footnote{Note that with wrong choice of the vielbeins this will
not be obvious.}. Note that \eqref{nonkah} is definitely not K\"ahler because the constraints \eqref{kahrest} are not satisfied.


Let's now consider supersymmetry. The interesting elements of the torsion classes that
are responsible for determining supersymmetry are the $\mathcal{W}_4$ and the $\mathcal{W}_5$ torsion classes. The ($\mathcal{W}_4, \mathcal{W}_5$) torsion classes are:
\bg\label{w44w55}
&&\mathcal{W}_4  =  {F_{3r} - \sqrt{F_1 F_2}\over 4 F_3}  + {F_{4r} -  \sqrt{F_1 F_2}\over 4 F_4} + \phi_r , \nonumber\\
&&{\rm Re}~\mathcal{W}_5  =  {F_{3r} \over 12 F_3} + {F_{4r} \over 12 F_4} + {F_{2r} - 2\sqrt{F_1F_2} \over 12 F_2} + 
{\phi_r\over 2} . \nd
Comparing the $\mathcal{W}_5$ torsion class with the constraint \eqref{bgcos}, and now assuming that the dilaton is constant, $\phi=\phi_0 $, then it is no surprise that we have:
\bg\label{jhotke}
{\rm Re}~\mathcal{W}_5 = 0 , \nd
and thus the supersymmetry condition for the background \eqref{kunku} is simply:
\bg\label{kunsusy}
2\mathcal{W}_4 + {\rm Re}~\mathcal{W}_5 = 2 \mathcal{W}_4 = {1\over 2} \left({F_{3r} - \sqrt{F_1 F_2}\over F_3}  + {F_{4r} 
-  \sqrt{F_1 F_2}\over F_4}\right) = 0 , \nd  
which is precisely the susy condition that we had in \eqref{ghonti} for the constant dilaton case! Note that supersymmetry is unbroken as long as \eqref{kunsusy} vanishes up to a total derivative, 
since such a term can be absorbed as a rescaling of the metric \cite{Lopes Cardoso:2002hd}, and hence supersymmetry will be unbroken even in the varying dilaton case.

%
%

So far we managed to determine all the torsion classes except ${\cal W}_3$. 
The value of $\mathcal{W}_3$ can now be directly read off from the ($2, 1$) piece of $dJ$, namely:
\begin{equation}
[dJ] ^{(2,1)} = \left[  J \wedge \mathcal{W}_4\right]^{(2,1)} + \mathcal{W}_3  =\mathcal{W}_3 ,
\end{equation}
where the second equality is due to $\mathcal{W}_4 =0$, as per equation \eqref{kunsusy}. It then follows from \eqref{dj} that $\mathcal{W}_3$ is non-zero, which one would expect since $\mathcal{H}$ in \eqref{kunku} is non-vanishing.

Before we move on to the dualized IIB background, lets collect our current results for the torsion classes before any dualities for the special case 
where the background dilaton has no profile:
\begin{equation}
\mathcal{W}_1 = \mathcal{W}_2 = \mathcal{W}_4  = \mbox{Re}~ \mathcal{W}_5 = 0, \;\;\;\;\; \mathcal{W}_{3} \neq 0, \;\;\;\;\; 
 2\mathcal{W}_4 + {\rm Re}~\mathcal{W}_5=0 . 
\end{equation}
Therefore the original background \eqref{nonkah}, with a constant dilaton profile,
is a supersymmetric non-K\"ahler {\it special-Hermitian manifold}, i.e. a complex manifold with a closed holomorphic (3, 0) form and torsion determined only by the ${\cal W}_3$ class.

%

\subsection{Type IIB background after duality \label{aftdual}}

So far we have seen how the type IIB background \eqref{kunku} and \eqref{nonkah} with ${\cal H}$ torsion can be duality chased to another type IIB 
background, now with both ${\cal H}_3$ and ${\cal F}_3$ three-form fluxes. We will now see that the type IIB background after duality can have an 
integrable complex structure provided the original type IIB background before duality chasing is a special-Hermitian manifold: a manifold with 
constant dilaton profile and torsion only in ${\cal W}_3$ class. In this case, the precise background is:
\bg\label{kolaram}
&& ds^2 = {1\over \sqrt{H}} ds^2_{0123} + \sqrt{H} ds^2_6\nonumber\\
&& {\cal F}_3 = -{\rm cosh}~\beta ~e_{\psi}\wedge (e_{\theta_1} \wedge e_{\phi_1} + g_1~ e_{\theta_2} \wedge e_{\phi_2})\nonumber\\
&& {\cal H}_3 = -e^{-2\phi} {\rm sinh}~\beta \sqrt{F_1\over F_2} ~e_r \wedge (e_{\theta_1} \wedge e_{\phi_1}  + g_1~ e_{\theta_2} \wedge e_{\phi_2})\nd
where $H$ and $\Delta$ are both constants because the dilaton has no profile. For the case where $e^\phi = 1$, the background 
\eqref{kolaram} has only delocalized sources with:
\bg\label{hdelta} \Delta = 0, ~~~~~~ H = 1.\nd
One way to study complexity of the dual to above background is to compute the holomorphic (3, 0) form $\Omega$ using the vielbeins \eqref{c1fom}. 
We can define two functions ${\cal C}(r)$ and ${\cal D}(r)$ similar to the ones defined earlier in \eqref{calab}:
\bg\label{calcd}
{\cal C}(r) \equiv H^{3/4}\sqrt{F_1 F_3 F_4}, ~~~~~~~{\cal D}(r) \equiv \gamma H^{3/4}\sqrt{F_2 F_3 F_4} , \nd
such that the condition $d\Omega=0$ is precisely \eqref{condu}, but with ${\cal A}$ and ${\cal B}$ are replaced by ${\cal C}$ and ${\cal D}$ respectively. In terms of the
warp factors appearing \eqref{iibform}, $d\Omega$ turns out to be:
\bg\label{dela}
d\Omega = {F_{3r}\over F_3} + {F_{4r}\over F_4} + {\gamma F_{2r} - 2 \sqrt{F_1F_2}\over \gamma F_2} 
+ {2\gamma_r\over \gamma} + {3\over 2}\cdot {H_r\over H}. \nd  
For the special case of constant dilaton,  we have ${\cal A} = {\cal C}$ and ${\cal B}= {\cal D}$, and hence $d\Omega=0$. Similarly, $dJ$ has only (2,1) and (1,2) components, and hence the dualized manifold is complex.

 However, comparing \eqref{dela} with \eqref{w44w55}, we note that an equation like \eqref{photka} cannot generically be satisfied, i.e. for $\phi=\phi(r)$, unless ${\cal W}_2$ is switched on. Thus the type IIB
manifold in \eqref{iibform} is in general a non-complex non-K\"ahler manifold. This should not be a surprise: under duality transformations a complex manifold can
become a non-complex one.

The solution \eqref{kolaram} with \eqref{hdelta} is an interesting example with non-K\"ahler resolved confold background, and probably the {\it simplest} non-trivial extension of the 
well known Calabi-Yau case. However as we show below, this is not the only one. There are numerous choices of non-K\"ahler metric on a resolved conifold possible
if we allow for non-trivial dilaton profile. In fact such examples will have an added advantage: we will be able to argue for localized sources.

With this in mind, let us now assume that the original background \eqref{nonkah} is a complex non-K\"ahler manifold but not of the special-Hermitian kind, i.e we allow a
dilaton profile in \eqref{kunku}. 
The question now is: under the duality transformation 
that converted \eqref{kunku} to \eqref{iibform}, is the 
six-dimensional manifold in \eqref{iibform} also a complex non-K\"ahler manifold? 

To be more specific, we will take a concrete example motivated by \cite{pandoz}, but with varying dilaton. We will assume without loss of generality a before-duality metric 
of the form:
\bg\label{dhonsho}
ds^2 = ds^2_{0123} + {e^{2\phi}\over 2}\left[{e^2_r\over F(r)} + r^2F(r) e_\psi^2
+ {1\over 2}r^2(e_{\theta_2}^2 + e_{\phi_2}^2) + {1\over 2}(r^2 + 4a^2 e^{-2\phi})(e_{\theta_1}^2 + e_{\phi_1}^2)\right], \nonumber\\  \nd
where $F(r)$ is a function of $r$ and $e^\phi$ is the background dilaton. The above metric 
clearly falls in the class of metrics \eqref{kunku} with D5-brane wrapping the resolved two-sphere parametrized by ($\theta_1, \phi_1$). 
Note that in the language of \eqref{nonkah} the constant resolution parameter $a^2$ is defined as:
\bg\label{lesfis}
F_3(r) - F_4(r) = a^2 . \nd
The internal manifold in \eqref{dhonsho} is a complex non-K\"ahler manifold, and the condition for supersymmetry as before will 
become\footnote{Note that our convention differs from \cite{Lopes Cardoso:2002hd} as well as \cite{lapan}. In the latter (${\cal W}_4, {\cal W}_5$) as computed therein are equated to 
($d\phi/2, d\phi$) respectively. See also footnotes 11 and 15 in \cite{lapan}.}:
\bg\label{cforsusy}
{\cal W}_4 = d\phi, ~~~~~~~ {\rm Re}~{\cal W}_5 = - 2 d\phi , \nd
with non-trivial dilaton and upto ${\cal O}(a^2)$ corrections to RHS of \eqref{cforsusy}. 
Thus the internal manifold is no longer a special-Hermitian manifold, and it is easy to see 
using \eqref{w44w55} that the first condition on 
${\cal W}_4$ is satisfied up to terms proportional to ${\cal O}(a^2)$: 
\bg\label{murgi}
{\cal W}_4 = \left({r^2 + 2a^2 e^{-2\phi} \over r^2 + 4a^2 e^{-2\phi}}\right)\phi_r = \phi_r -\left({2\phi_r e^{-2\phi}\over r^2}\right)a^2 + 
{\cal O}(a^4). \nd
The second condtion in \eqref{w44w55} is more non-trivial and it leads to the following constraint on dilaton $\phi$ and 
the warp factor $F$:
\bg\label{thchini}
r{d\phi\over dr} + {r\over 30} {d\over dr}\left({\rm log}~F\right) + {1\over 5}\left(1 - {1\over 3F}\right) + {\cal O}(a^2) = 0. \nd
Once we dualize this background to generate the NS and RR three-form fluxes, the background takes the form \eqref{iibform}, but now the NS and RR three-form
fluxes are simpler:
\bg\label{altex}
&&{\cal H}_3 = -2 a^2 e^{-2\phi} \phi_r ~{\rm sinh}~\beta~ e_r \wedge e_{\theta_1}\wedge e_{\phi_1} ,\nonumber\\
&& {\cal F}_3 = -\left({2 a^2 r^3 F \phi_r \over r^2 + 4a^2 e^{-2\phi}}\right) ~{\rm cosh}~\beta~ e_\psi \wedge e_{\theta_2}\wedge e_{\phi_2} , \nd
where the non-vanishing of $d{\cal F}_3$ clearly reflects sources to be along the right directions. Note that the way we supported the D5-brane is via 
non-K\"ahlerity generated by varying resolution parameter $a^2 e^{-2\phi}$. Thus when $a$ vanishes, we have no D5-brane.
The metric, of course still takes the form \eqref{kolaram}, but now both $H$ and $\Delta$ are non-trivial functions of the radial coordinate and the resolution parameter. Finally, the type IIA gauge field coming from the 
T-dual D6-brane along ($\theta_1, \phi_1, \psi$) is now given by:
\bg\label{biratpa}
{\cal F}_2 = -2 a^2 r F \phi_r ~{\rm cosh}~\beta~e_{\theta_2} \wedge e_{\phi_2} + {\cal O}(a^4) \equiv 
\widetilde{g}_1~ {\rm cosh}~\beta~e_{\theta_2}\wedge e_{\phi_2} , \nd
{}from which the corresponding gauge field can be easily determined as before.

It is clear that the dualized manifold, with warp-factor $H$ instead of $e^{2\phi}$, 
cannot be K\"ahler, however it remains to be checked if complexity is maintained. To check this, let us assume that the complex vielbeins are of the
form \eqref{c1fom} with an almost complex structure $\gamma$. This complex structure cannot be integrable because if it were then \eqref{choddal}
will have both (2, 1) and (1, 2) components implying breaking of supersymmetry. However the fluxes in \eqref{altex} are explicitly supersymmetric because we have used 
\eqref{cforsusy} to compute them. The resolution of this is that the manifold after duality does not have an integrable complex structure,
at least using the choice of vielbeins that we have taken. 


\section{Branes Lifted to M-theory: Geometry and Harmonic Forms \label{sec4}}

We would now like to study these solutions from M-theory. However, before we do so, we will consider some simple examples. Let us begin with a very basic scenario of the lift of a D6-brane to M-theory. This is typically given by a
Taub-NUT space with the following metric:
\begin{equation}\label{sixb}
ds^2 = ds^2_{012.....6} + H_\alpha(dr^2 + r^2 d\Omega^2) + H_\alpha^{-1} (d\psi + \alpha \cdot{\rm cos}~\theta d\phi)^2 ,
\end{equation}
where $H_\alpha$ is the standard harmonic function whose value is given by $H_\alpha = 1 + {\alpha\over r}$ with $\alpha$ being a 
constant. Note that if \eqref{sixb} is the localized solution 
obtained from T-dualizing a D5-brane to a D6-brane (the case that we are interested in), 
then it has the 
correct warp factor (or, in this language, the correct harmonic function). The constant $\alpha$ is defined as:
\begin{equation}
\alpha = {g_s l_s^2\over 2R} ,
\end{equation}
where $R$ is the radius of the single-centred Taub-NUT space at $r \to \infty$ i.e at spatial infinity. 

The D6-brane world-volume theory is encoded in the M-theory geometry via a normalizable two-form $\omega$ \cite{imamura, ashoke, Sen:1994yi, robbins}. In simple examples where the M-theory geometry is a 4d Taub-Nut space in a M-theory fourfold, the task of finding a normalizable two-form is simplified by the fact that the space of 2-forms on a 4d space decomposes into two subspaces, corresponding to self-dual and anti-self dual forms. Hence it suffices to search for such a form, and then test for normalizability. This method has been applied in the past, see for example \cite{robbins}. We will compute this form explicitly, both in the absence and presence of fluxes.

In the presence of G-fluxes, the background metric \eqref{sixb} changes. For a generic choice of G-fluxes, the change in 
the metric components can be determined by solving the EOMs. This is in general hard as the background 
EOMs are highly non-trivial (we will discuss this soon). There is, however, a simple {\it trick} by which 
one may determine certain aspects of the change in metric, using the type IIB D5-brane. The idea is to 
take the five-brane solution and {\it twist} the background solution along the orthogonal direction of the 
five brane. This twist is effectively performed for the case where the D5-brane solution is delocalized
along the orthogonal direction. We then T-dualize the twisted solution along the twist-direction and 
lift the solution to M-theory. In M-theory we get the required deformed Taub-NUT metric in the presence
of certain components of the G-flux. 

To define the twist properly we need to analyze the asymptotic behavior of the G-flux. We use ${\cal G}_4 = d{\cal C}_3$ 
to define the twist as:
\begin{equation}
{\cal C}_{z_1 z_2\psi}(r \to \infty) - {\cal C}_{z_1 z_2\psi}(r \to 0) \equiv {\rm tan}~\alpha , 
\end{equation}
where the directions ($z_1, z_2$) are related to the directions ($x^5, x^6$) in \eqref{sixb} as:
\begin{equation}
\left(\begin{matrix} x^5 \\ x^6 \end{matrix}\right) = \left(\begin{matrix}{\rm sec}~\alpha & ~~~R~ {\rm sin}~\alpha \\ 0 & ~~~R~ {\rm cos}~\alpha \end{matrix}\right) 
\left(\begin{matrix} z_1 \\ z_2 \end{matrix}\right) . 
\end{equation}
The G-flux associated with this twist can be expressed
in the following way:
\bg\label{gfla}
{\cal G}_4 = && {2R' {\rm sin}~\alpha\over (2r + R' {\rm cos}~\alpha)^2} ~ dr \wedge dz_1 \wedge dz_2 \wedge 
\left(d\psi + {R'\over 2} {\rm cos}~\theta ~d\phi\right) \nonumber\\
&& ~~~~~~~~~- {r R' {\rm tan}~\alpha {\rm sin}~\theta \over 2r + R'{\rm cos}~\alpha}
dz_1 \wedge dz_2 \wedge d\theta \wedge d\phi , 
\nd 
with $R' = {g_s l_s^2\over R}$ as the new scale, and we see clearly that the flux vanishes in the limit $\alpha \rightarrow 0$. Note that this G-flux when reduced to type IIA will give rise to the necessary 
$B_{NS}$ field. The M-theory metric that solves EOM with the flux choice \eqref{gfla} can be expressed as:
\bg\label{meturic}
ds^2 = && \left({2r {\rm cos}~\alpha + R' {\rm cos}^2\alpha \over 2r {\rm cos}~\alpha + R'}\right)^{1/3} ds^2_{01234} + 
\left({2r {\rm cos}~\alpha + R' \over 2r {\rm cos}~\alpha + R' {\rm cos}^2\alpha}\right)^{2/3} ds^2_{z_1z_2} \nonumber\\
&& + ~{\left(2r {\rm cos}~\alpha + R'\right)^{2/3} \left(2r + R'{\rm cos}~\alpha\right)^{1/3}
\over 2r ~{\rm cos}^{2/3}\alpha} 
(dr^2 + 
r^2 d\Omega_2^2) \nonumber\\
&& + ~ {2r ~{\rm cos}^{1/3}\alpha \over 
\left(2r {\rm cos}~\alpha + R'\right)^{1/3} \left(2r + R'{\rm cos}~\alpha\right)^{2/3}} 
\left(d\psi + {R'\over 2} {\rm cos}~\theta ~d\phi\right)^2 , 
\nd 
where $R'$ is inversely proportional to $R$, the asymptotic radius of the Taub-NUT space, i.e
\bg R' \equiv {g_s l_s^2\over R} . \nd 
The M-theory metric that we are dealing with now has the following form:
\bg\label{juta}
ds^2 = && G_1(r) ds^2_{01234} + G_2(r) ds^2_{z_1 z_2} + G_3(r, \theta) dr^2 + G_4(r, \theta) d\theta^2 + G_5(r, \theta) d\phi^2 \nonumber\\
&&  ~~~~~~~~~~~~ + G_6(r, \theta) \left(d\psi + {1\over 2} R' {\rm cos}~\theta~
d\phi\right)^2  , 
 \nd
where  $G_i(r)$ are the warp factors that could, for example, be read from 
a variant of the metric \eqref{meturic}. The above metric \eqref{juta} could allow for wrapped D7-branes also.

Following \cite{robbins} we can construct the normalizable two-form $\omega$ by first defining a one-form $\zeta$ in the following way:
\bg\label{1form}
\zeta ~ \equiv g_1(r, \theta) \left(d\psi + {1\over 2} R' {\rm cos}~\theta~d\phi\right) + g_2(r, \theta) d\phi ,
\nd
and we define $\omega \equiv d \zeta$. If we now demand $\omega$ to be self-dual or anti-self-dual on the Taub-Nut space, i.e $\omega = \pm \ast_4 \omega$, and also normalizable,
then $g_1(r, \theta)$ and $g_2(r)$ must satisfy the following relations:
\bg\label{gvalue}
&& {\partial g_1\over \partial r} \sqrt{G_4 G_5\over G_3 G_6} =  \left(-{1\over 2} R' g_1 {\rm sin}~\theta + 
{\partial g_2 \over \partial\theta}\right) \nonumber\\
&& {\partial g_1 \over \partial\theta} \sqrt{G_3 G_5\over G_4 G_6} = - {\partial g_2 \over \partial r} , 
\nd
provided $\omega$
is self-dual (SD), i.e $\omega = \ast_4 \omega$. For $\omega$ anti-self-dual, it is not possible to find a 
normalizable harmonic two-form\footnote{This statement is dependent on the choice of veilbeins. For a different choice of veilbeins, it may be the anti-self-dual solution which is normalizable. }. 

A solution for the set of equations \eqref{gvalue} can be constructed in the following way. First let us assume that 
${\partial g_1 \over \partial \theta}$ is non-zero. For this case, we can have:
\bg\label{g2}
g_2(r, \theta) = g_2(\infty, \theta) + \int_r^\infty dr~ {\partial g_1\over \partial\theta} \sqrt{G_3G_5\over G_4 G_6} .
\nd
Interestingly if \eqref{g2} is {\it independent} of $\theta$, then $g_1$ takes the following form in terms of the warp factors $G_i$:
\bg\label{g1now}
g_1(r, \theta) = g_{0}~ {\rm exp}\left[- {R'\over 2} \int_r^\infty dr~{\rm sin}~\theta \sqrt{G_3 G_6\over G_4 G_5} \right] ,
\nd
which would happen if 
\bg\label{deriv}
{\partial g_2(\infty, \theta)\over \partial \theta} = - \int_r^\infty dr {\partial \over \partial \theta} 
\left({\partial g_1\over \partial\theta} \sqrt{G_3G_5\over G_4 G_6}\right) .
\nd
The above equation \eqref{deriv} looks highly constrained because the LHS is independent of the radial coordinate $r$ whereas the RHS 
could in principle depend on $r$ for more generic choices of the warp factors $G_i(r, \theta)$. Thus one would have to tread more 
carefully here. In the following let us take few examples to clarify the scenario.

\subsection{A warm-up example: Regular D6-brane \label{warm1}}

The simplest case of a regular D6-brane, i.e. a D6 brane without background flux, is easy. We will take the twist parameter $\alpha = 0$ in both \eqref{gfla} and \eqref{meturic} 
giving us vanishing G-flux. 
For this case we have:
\bg\label{reg7} 
&&G_3(r, \theta) = 1 + {R'\over 2r}, ~~~~~~ G_6(r, \theta) = \left(1 + {R'\over 2r}\right)^{-1} , \nonumber\\
&& G_4(r, \theta) = r^2 G_3(r), ~~~~~~ G_5(r, \theta) = r^2 {\rm sin}^2\theta~G_3(r) , \nd
as the background warp-factors. Note that this converts the metric \eqref{meturic} completely to a standard Taub-NUT space, and the 
differential equations \eqref{gvalue} to the following:
\bg\label{gvalnow}
&&r^2 {\rm sin}~\theta \left(1+{R'\over 2r}\right) {\partial g_1\over \partial r} = -{1\over 2} R' g_1 {\rm sin}~\theta + 
{\partial g_2 \over \partial \theta} , \nonumber\\
&& {\partial g_1 \over \partial \theta} \left(1+{R'\over 2r}\right) {\rm sin}~\theta = - {\partial g_2\over \partial r} .
\nd
The above two first-order equations give rise to the following second-order equation for $g_1(r, \theta)$ in terms of the variables 
appearing in them:
\bg\label{secor}
r^2 {\partial^2 g_1 \over \partial r^2} + 
2r {\partial g_1 \over \partial r} = -{\partial^2 g_1 \over \partial \theta^2} - 
{\partial g_1 \over \partial \theta} {\rm cot}~\theta  . \nd
It is interesting that the above differential equation is independent of $R'$, but this constant will appear when we fix the boundary conditions.
The above differential equation may be solved using separation of variables in $r$ and $\theta$ coordinates. This amounts to the 
assumption that $g_1(r, \theta) = g_a(r) g_b(\theta)$, leading us to the following equations for $g_a$ and $g_b$:
\bg\label{gagb}
&&r^2 {d^2g_a\over dr^2} + 2r {dg_a\over dr} = \lambda g_a , \nonumber\\
&& {1\over {\rm sin}~\theta} {d\over d\theta}\left({\rm sin}~\theta ~{dg_b\over d\theta}\right) = -\lambda g_b ,
\nd
where $\lambda$ is the eigenvalue. 

Let's first consider the zero-mode, i.e. $\lambda=0$, and we will come back to general $\lambda$ soon.  For $\lambda=0$ the second equation
in \eqref{gagb} gives us:
\bg\label{gb}
g_b(\theta) = c_0 + c_1 {\rm log}\left({\rm tan}~{\theta\over 2}\right) ,
\nd
where ($c_0, c_1$) are constants. The above solution blows up at $\theta = 0$ and so to avoid pathologies we can put $c_1 = 0$. Thus 
$g_b$ is just a constant and we can normalize this to $g_b = 1$. This way $g_1(r, \theta)$ is completely independent of the angular 
coordinate, and is given by:
\bg\label{ga}
g_1(r, \theta) \equiv g_a(r) = c_3\left(1 + {R'\over 2r}\right) , \nd
with $c_3$ the normalization constant. This can be seen from the fact that 
once $g_1$ is independent of $\theta$, the second equation in \eqref{gvalnow} implies that $g_2$ is independent of the radial coordinate $r$. Now plugging the first equation \eqref{gagb} in the first equation of 
\eqref{gvalnow} implies that:
\bg\label{g2eq} 
{1\over \sin~\theta}\cdot {\partial\over \partial r}\left({\partial g_2\over \partial \theta}\right) = 
\left(1+ {R'\over 2r}\right)\lambda g_1(r, \theta), \nd
and hence $\lambda = 0$ implies $g_2$ is independent of $\theta$ also! 
This means the $g_2$ part in \eqref{1form} is a total derivative and is therefore 
trivial in cohomology. Finally, the normalizable one-form $\zeta$ is given by:
\bg\label{zetbeta}
\zeta = c_3 \left(1 + {R'\over 2r}\right) \left(d\psi + {1\over 2} R'\cos~\theta~d\phi\right).\nd   
So far our discussion was related to the zero mode only. What about other values of $\lambda$? To study this, note that the second equation
in \eqref{gagb} is a Legendre equation in $\theta$ variable, implying that:
\bg\label{legen}
g_b(\theta) = P_n({\cos}~\theta), ~~~~~~~~ \lambda = n(n+1) . 
\nd
Using the above value for $\lambda$, we can formulate the first equation in \eqref{gagb}:
\bg\label{1eq}
{d\over dr}\left(r^2 {dg_a\over dr}\right) = n(n+1) g_a , \nd
as the large $x$ limit of the Legendre equation for $P_n(x)$. This means that the large $r$ limit for $g_a(r)$ will 
be given by:
\bg\label{largeg}
g_a(r) \approx 2^n \left(\begin{matrix} {2n-1\over 2} \\ n \end{matrix}\right) r^n , \nd 
for $\lambda = n(n+1)$. These functions are clearly non-normalizable, and therefore the corresponding two-form $\omega$ will not be 
a normalizable harmonic form. Thus the {\it only} choice is for $n = 0$ or $\lambda = 0$, which is \eqref{zetbeta}, in accordance with the 
previous results that a single-centered Taub-NUT space has a unique normalizable harmonic two-form  \cite{imamura, ashoke, Sen:1994yi, robbins}.

\subsection{Another warm-up example: D6-brane with background fluxes \label{warm2}}

Let us now discuss the case of the D6-brane in the presence of fluxes. In M-theory this will be the background studied in 
\eqref{meturic} and \eqref{gfla}. The warp factors are given by:
\bg\label{gonu}
&& G_4 = r^2 G_3, ~~~~~~~~~ G_5 = r^2 {\sin}^2\theta~G_3\nonumber\\
&&G_3 = {\left(2r {\cos}~\alpha + R'\right)^{2/3}\left(2r + R'{\cos}~\alpha\right)^{1/3}\over 2r {\cos}^{2/3}\alpha}\nonumber\\
&&G_6 = {2r {\cos}^{1/3}\alpha \over \left(2r {\cos}~\alpha + R'\right)^{1/3}\left(2r + R'{\cos}~\alpha\right)^{2/3}}  , 
\nd
where $\alpha$ is the required twist parameter. The equation for $g_1(r, \theta)$ follows from the same procedure outlined in
\eqref{gvalue}. Thus as before, decomposing $g_1(r, \theta) = g_a(r) g_b(\theta)$, the equation for $g_a$ is given by:
\bg\label{dhong}
\lambda g_a = && r^2 {d^2 g_a\over dr^2} + 2r ~{dg_a\over dr}\\
&& - {R'\over 4} { d g_a\over dr} 
{1\over \sqrt{1 + {R'\over 2r}\left(\cos~\alpha + {\rm sec}~\alpha\right) + \left({R'\over 2r}\right)^2}}
\left[{\cos~\alpha + {\rm sec}~\alpha + {R'\over r} \over \sqrt{1 + {R'\over 2r}\left(\cos~\alpha + {\rm sec}~\alpha\right) + \left({R'\over 2r}\right)^2}} - 2\right] , \nonumber
\nd
and the equation for $g_b(\theta)$ is similar to the second equation in \eqref{gagb}. This means we can again take the $\lambda = 0$ solution,
again implying the $\theta$ independence of $g_1$. Additionally, using arguments mentioned earlier, $g_2(r, \theta)$ is 
just a constant as before. Therefore the solution for $g_a(r)$ or $g_1(r, \theta)$ is then given by:
\bg\label{churi}
g_1(r, \theta) \equiv g_a(r) = {c_3\over 4} \left[\cos~\alpha + {\rm sec}~\alpha + {R'\over r} + 2\sqrt{1 + 
{R'\over 2r}\left(\cos~\alpha + {\rm sec}~\alpha\right) + \left({R'\over 2r}\right)^2}\right] , \nonumber\\  
\nd
where $c_3$ is the same constant that appeared before. Its comforting to see that $\alpha = 0$ limit reproduces the result for the vanilla (i.e. flux less)
Taub-NUT. Finally, in the limit of large $r$, or more concretely, for 
\bg\label{larr}
r ~>> ~ {R'\over 8}\left(\cos~\alpha + {\rm sec}~\alpha - 2\right) , \nd
normalizable solution only exists for $\lambda = 0$. Thus, expectedly, there exists a unique normalizable harmonic form for the Taub-NUT
background with G-flux.

\section{M-theory Lift of the Five-brane on a Warped Resolved Conifold \label{sec5}}

Our next example is a more complicated one: a D5-brane wrapped on a warped resolved conifold, or D6-brane embedded in a related
background. We will continue using the M-theory description as the analysis will be easier to perform. The wrapped D5-branes are converted to the 
D6-branes which are then lifted to M-theory. As we discussed earlier, the D5-brane should be delocalized along the orthogonal T-duality direction.
 
The conifold and its cousins, the resolved 
and deformed conifolds, can be Calabi-Yau spaces if one allows Ricci flat metrics on them. Generically, however, they will allow non-K\"ahler metrics.
One may wrap branes on appropriate cycles of the conifolds and get the 
corresponding world-volume dynamics and effective theory on the non-compact directions of the branes. 

As an example of lifting a conifold to M-theory, let us consider the case of D6-branes wrapped on the three-cycle of a deformed conifold. 
We reach this configuration by taking the SYZ \cite{syz} mirror of the wrapped D5-brane on a resolved conifold.
How do the dynamics look from M-theory? This is not a new question and has been addressed in 
recent papers like \cite{fangpaul}, where the M-theory lift for a special case, with appropriate G-fluxes, is expressed as:
\bg\label{mliftnow}
ds^2_{11}=&&e^{-{2\phi\over 3}}\Bigg\{F_0ds_{0123}^2+F_1 dr^2 + {\alpha F_2
\over \Delta_1 \Delta_2} \Big[d\psi -b_{\psi r}dr - b_{\psi\theta_2} d\theta_2\nonumber\\
&&+ \Delta_1 {\rm cos}~\theta_1 \Big(d\phi_1 - b_{\phi_1\theta_1}
d\theta_1-b_{\phi_1 r}dr\Big)+ \Delta_2 {\rm cos}~\theta_2 {\rm cos}~\psi_0
\Big(d\phi_2 - b_{\phi_2\theta_2} d\theta_2-b_{\phi_2
r}dr\Big)\Big]^2\nonumber\\
&& + \alpha {\cal E}_1\Big[d\theta_1^2 + \Big(d\phi_1 -
b_{\phi_1\theta_1} d\theta_1-b_{\phi_1 r}dr\Big)^2\Big] \nonumber\\
&&+ \alpha {\cal E}_2\Big[d\theta_2^2 + \Big(d\phi_2 -
b_{\phi_2\theta_2} d\theta_2-b_{\phi_2
r}dr\Big)^2\Big]\nonumber\\
&& + 2\alpha {\cal E}_3~{\rm cos}~\psi_0\Big[d\theta_1 d\theta_2 - \Big(d\phi_1
- b_{\phi_1\theta_1} d\theta_1-b_{\phi_1 r}dr\Big)\Big(d\phi_2 -
b_{\phi_2\theta_2} d\theta_2-b_{\phi_2
r}dr\Big)\Big]\nonumber\\
&& + 2\alpha {\cal E}_3~{\rm sin}~\psi_0\Big[\Big(d\phi_1- b_{\phi_1\theta_1}
d\theta_1-b_{\phi_1 r}dr\Big) d\theta_2
+ \Big(d\phi_2 - b_{\phi_2\theta_2} d\theta_2-b_{\phi_2r}dr\Big)d\theta_1\Big]\Bigg\}\nonumber\\
&&+e^{4\phi\over 3}\Big[dx_{11}+{A}_{\phi_1}d\phi_1+{A}_{\phi_2}d\phi_2+
{A}_{\theta_1}d\theta_1
+{A}_{\theta_2}d\theta_2+{A}_rdr\Big]^2 , 
\nd
where $F_i$ are the warp factors such that $F_0 = F_0(r, \theta_1, \theta_2)$ and $F_k = F_k(r)$ for $k \ne 0$, and ($\theta_i, \phi_i$)
with $\psi$ are the usual deformed conifold coordinates \cite{candelas}. 
In fact the $F_i(r)$ are exactly the same $F_i$ described in \eqref{nonkah} and the metric components used in \eqref{mliftnow} are: 
\bg\label{metko}
&& {\cal E}_1 = F_2 \cos^2\theta_2 + F_4 \sin^2\theta_2, ~~~ {\cal E}_2 = F_2 \cos^2\theta_1 + F_3 \sin^2\theta_1 , \nonumber\\
&& \alpha^{-1} = {F_3 F_4 {\rm sin}^2\theta_1 {\rm sin}^2\theta_2 + F_2 F_4 {\rm cos}^2\theta_1{\rm sin}^2\theta_2 
+ F_2 F_3 {\rm sin}^2\theta_1 {\rm cos}^2\theta_2} , \nonumber\\
&&{\cal E}_3 = F_2 \cos~\theta_1~\cos~\theta_2, ~~~ \Delta_1 =  \alpha F_2 F_4 {\rm sin}^2 \theta_2, 
~~~ \Delta_2  = \alpha F_2 F_3 {\rm sin}^2 \theta_1 . \nd
The ($b_{mn}, {\cal E}_i, \phi$) are 
respectively the components of the $B_{NS}$ field, 
the metric and the dilaton in the
dual type IIB side and $A_M$ are the type IIA $U(1)$ gauge field components. We have also defined $\psi_0 = \langle \psi\rangle$ and 
$\Delta_i$ to be the warp factors that depend on all the above parameters. For details the readers may refer to \cite{fangpaul}.

The wrapped D6-branes in type IIA are oriented along ($\theta_2, \phi_1$) and $\psi$ in the internal space, and spread along 
spacetime directions $x^{0,1,2,3}$. In M-theory, the deformed Taub-NUT space will take the following form:
\bg\label{orTN}
ds^2 = e^{-{2\phi/ 3}}\left[F_1 dr^2 + A \big|d\theta_1 + \tau d\phi_2\big|^2\right] +
e^{4\phi/3}\left(dx_{11} + A_{\phi_2} d\phi_2 + A_{\theta_1} d\theta_1\right)^2
, \nd
where we have restricted ourselves to the case with $b_{rm} = 0$; and switched on a complex structure $\tau$ defined as:
\bg\label{chuk}
&& \big|\tau \big|^2 = \alpha A^{-1}\left[{\cal E}_2 + F_2 \Delta_2 \Delta_1 ^{-1} \cos^2\theta_2 ~\cos^2 ~\psi_0\right]\nonumber\\
&& {\rm Re}~\tau = \alpha A^{-1}\left[{\cal E}_3 (\sin~\psi_0 - b~\cos~\psi_0) + b~F_2 \cos~\theta_1 \cos~\theta_2~\cos~\psi_0\right] ,
\nd
where $b = b_{\theta_1 \phi_1}$ and the coefficient $A$ appearing in \eqref{orTN} and \eqref{chuk} is defined as:
\bg\label{Adof}
A \equiv \alpha\left[F_2 \Delta_1 \Delta^{-1}_2 b^2 \cos^2\theta_1 + {\cal E}_1 (1 + b^2)\right] . \nd 
Hence lift of the D6 brane wrapped on a deformed conifold is not of a simple Taub-NUT form, and additionally there are cross-terms that would
make the analysis of the harmonic form highly non-trivial. 

Let us now return to the problem at hand: D5 brane wrapped on the resolved conifold. We have already developed the details of the solution, including its M-theory lift, so our
next step should be obvious. However new subtleties arise because of two reasons: one, 
the D5-brane is no longer on a flat background as it wraps a two-cycle, 
and two, due to the existence of the non-trivial functions $g_1(r)$ in the RR gauge field  \eqref{jomadar} and $\widetilde{g}_1(r)$ in 
\eqref{biratpa}. 

First let us consider the wrapped D5-brane in the delocalized limit. 
A standard T-duality along an orthogonal direction should convert this to a wrapped D6-brane. The $C_7$ source charge of the D6-brane decomposes
in the following way:
\bg\label{c7}
C_7(\overrightarrow{\bf x}, \psi, \theta_1, \phi_1) ~ = ~ 
C_5(\overrightarrow{\bf x}, \psi) \wedge \left({e_{\theta_1} \wedge e_{\phi_1} \over \sqrt{V_2}}\right)\nd
where $V_2$ is the volume of the two-sphere that is being wrapped by the D6-brane and whose cohomology is represented by the term in the 
bracket\footnote{The representative of second cohomology for a two-cycle of a conifold is $e_{\theta_1} \wedge e_{\phi_1} -
e_{\theta_2} \wedge e_{\phi_2}$ as both ${\bf P}^1$ vanish at the origin \cite{DM2}. 
For resolved conifold we will take \eqref{c7}, as
geometrically the D5-brane wraps a two-sphere parametrized by ($\theta_1, \phi_1$). This makes sense as one of the sphere
will be of vanishing size at $r = 0$.}. 
In the limit where 
the size of the two-sphere is vanishing (i.e for the T-dual conifold), the term in the bracket in \eqref{c7} will behave as a delta-function, and
consequently $C_7$ will decompose as $C_5$ i.e as a D4-brane. It will take infinite energy to excite any mode along the 
directions of the vanishing two-sphere, and 
therefore for all practical purpose a T-dual of the wrapped D5-brane on a conifold will be a D4-brane stretched along $\psi$ direction. This is 
of course the main content of \cite{uranga, DM, DM2}. Similarly if the wrapped two-sphere is of {\it finite} size, i.e the D5-brane
wraps the two-cycle of a resolved conifold, then at energy lower than the inverse size of the two-sphere i.e 
$1/\sqrt{a^2}$, the T-dual will effectively behave again as a 
D4-brane \cite{tatar1, tatar2}. Once the energy is bigger than this bound, or once $a^2 >> \alpha'$ i.e 
the size of the two-cycle much bigger than the string scale, then
the intermediate energy physics will probe the full D6-brane. Our analysis in this paper will be related to this case only, i.e we will explore
the classical dynamics of a wrapped D6-brane. 

To analyze the second subtlety, we need a more detailed study.
To start, let us consider the special-Hermitian case studied in \eqref{kolaram} with background values \eqref{hdelta}. With some minor alterations
we can extend our technique to encompass \eqref{dhonsho}, as we will discuss later. 
We also have a non-trivial five-form 
\eqref{5form} but a  
trivial dilaton $\phi_B = -\phi_0$. Note that ${\cal H}_3$ is 
closed but ${\cal F}_3$ is not, as expected. In fact:
\bg\label{dcal}
d{\cal F}_3 = -{\rm cosh}~\beta~g_1'(r) ~e_r \wedge e_{\psi} \wedge e_{\theta_2} \wedge e_{\phi_2} + {\rm cosh}~\beta~
e_{\theta_1} \wedge e_{\phi_1} \wedge e_{\theta_2} \wedge e_{\phi_2} , \nd
where the first term is the {\it delocalized} source term for the wrapped five-brane along the ($\theta_1, \phi_1$) sphere and stretched along the Minkowski directions
$x^{0, 1, 2, 3}$. The second term is a form that vanishes at every point on the two-sphere ($\theta_1, \phi_1$). This is more obvious in the T-dual 
type IIA description where the source equation \eqref{sourceq} is precisely the first term of \eqref{dcal}. 

We also need to determine the type IIA gauge field from the field strength \eqref{jomadar}. Since ${\cal F}_2$ is not closed, as per equation \eqref{sourceq}, and depending on
how we distribute charges in \eqref{jomadar},
we can rewrite ${\cal F}_2$ in at least two possible ways, i.e as: 
\bg\label{calf2}
{\cal F}_2 = d{\cal A}_1 - {\rm cosh}~\beta ~{\rm cot}~\theta_2~g_1'(r)~e_r \wedge e_{\phi_2}\nd
with components along ($r, \phi_2$), or as:
\bg\label{calf3}
{\cal F}_2 = d{\cal A}_1 + (1-g_1) {\rm cosh}~\beta~e_{\theta_2} \wedge e_{\phi_2}\nd
with components along ($\theta_2, \phi_2$). For the distribution \eqref{calf2}, 
the one-form gauge-field ${\cal A}_1$ is given by:
\bg\label{cala1}
{\cal A}_1 = {\rm cosh}~\beta\left(\cos~\theta_1 d\phi_1 + g_1(r) \cos~\theta_2 d\phi_2\right), \nd
whereas for the distribution \eqref{calf3} the one-form gauge-field ${\cal A}_1$ is given by:
\bg\label{kalabeta}
{\cal A}_2 = {\rm cosh}~\beta\left(\cos~\theta_1 d\phi_1 + \cos~\theta_2 d\phi_2\right). \nd
Observe that the function $g_1$ does not appear in the definition of the gauge-field in \eqref{kalabeta}. Finally note 
that there is a possible variant of \eqref{cala1} in which the $g_1$ function is shifted in the following way: 
\bg\label{kalasika}
{\cal A}_3 &=& {\rm cosh}~\beta\left(g_1\cos~\theta_1 d\phi_1 + \cos~\theta_2 d\phi_2\right)\nonumber\\ 
&=& {\cal A}_1 + {\rm cosh}~\beta (1-g_1) \left(\cos~\theta_2~d\phi_2 
- \cos~\theta_1~d\phi_1\right), \nd
where the angular piece in the second equality, appearing as a difference between two one-forms, will be the representative of the second cohomology under $d$-action for a conifold.
  
The output of our discussion above reveals that 
at every point ($\theta_1, \phi_1$) there is a deformed Taub-NUT space parametrized by the warp factors ($G_3, G_4, G_5, G_6$) in \eqref{hingsro2} with
the following ansatze for the one-form:
\bg\label{chukkabeta}
\zeta = g_2(r, \theta_2) \Big[d\Psi + g_1(r) \cos~\theta_2 d\phi_2\Big] , \nd
where $d\Psi = dx_{11}/{\rm cosh}~\beta$, and 
we have used $g_1(r)$ to represent all the choices \eqref{cala1}, \eqref{kalabeta} and \eqref{kalasika}, with the understanding that for the latter 
two choices $g_1 \to 1$ in \eqref{chukkabeta}. 
Self-duality and anti self-duality of the two-form $\omega \equiv d\zeta$ then imply the following two conditions on the coefficient 
$g_2$, using $G_5 = G_4 \sin^2\theta_2$:
\bg\label{condg2} 
{1\over g_2}{\partial g_2\over \partial r} = \pm g_1 {\sqrt{G_3G_6}\over G_4}, 
~~~~~~~{1\over g_2 ~{\rm cot}~\theta_2} {\partial g_2\over \partial \theta_2} = \pm {\partial g_1\over \partial r}\sqrt{G_6\over G_3} . \nd
To solve the set of equations \eqref{condg2}, we will use the usual separation of variables trick, defining:
\bg\label{ghatot}
g_2(r, \theta_2) \equiv \Lambda_1(r) \Lambda_2(\theta_2) . \nd
It is easy to solve for $\Lambda_1(r)$ once we plug-in \eqref{ghatot} in \eqref{condg2}. Using the warp factors $G_i$ in \eqref{hingsro2}, $\Lambda_1(r)$ 
is given by:
\bg\label{lattobaaz}
\Lambda_1(r) = \Lambda_0~ {\rm exp}\left[-\int^r dr~\vert g_1(r)\vert {\sqrt{G_3 G_6}\over G_4}\right] , \nd
with $\Lambda_0$ a constant and we have chosen the \emph{anti}-self-duality condition on $\omega$, to allow for normalizability.

So far the analysis has followed more or less the path laid out in the previous section. However, a subtlety appears once we study the $\Lambda_2$ equation. This is
given by:
\bg\label{subtle}
{1\over \Lambda_2} {d\Lambda_2\over d\theta_2} = - {\partial \vert g_1 \vert \over \partial r}\sqrt{G_6\over G_3} ~{\rm cot}~\theta_2 , \nd
where we see that the separation of variables trick has failed because of the $r$ dependent terms. Clearly this problem disappears when $g_1$ is a 
constant for the choices \eqref{kalabeta} and \eqref{kalasika}.

The simplest way out of this conundrum would be to evaluate $\partial^2 g_2/\partial r \partial \theta_2$ for both the equations in \eqref{condg2} and
compare. This immediately leads us to the following constraint on $g_1(r)$ of the form:
\bg\label{addcons}
{\partial \vert g_1\vert  \over \partial r} = \sqrt{G_3\over G_6} =   e^\phi \sqrt{H F_1 F_2} \nd
The above constraint is in {\it addition} to the earlier constraints \eqref{ghonti}, \eqref{whatif} and \eqref{lerae} imposed by supersymmetry. 

We are now ready to determine the one-form $\zeta$, given in \eqref{chukkabeta}, from the information we have gathered thus far. Using \eqref{lattobaaz}, 
\eqref{subtle} and the constraint \eqref{addcons}, the one-form is:
\bg\label{haro}
\zeta = g_0~ \sin~\theta_2~{\rm exp}\Bigg[-\int^r dr~{\vert g_1(r) \vert e^{-5\phi/3}} \sqrt{F_1\over \left(e^{2\phi/3}+ \Delta\right) F_2 F_4^2}\Bigg]
 \left(d\Psi + g_1~\cos~\theta_2 ~d\phi_2\right) \nonumber\\ , \nd  
where $g_0$ is a normalization constant, and $\Delta$ is defined in \eqref{linfri}.
Note that the normalizable form $\omega = d\zeta$ would work for any 
non-K\"ahler resolved conifold background whose warp factors appearing in the metric ansatze \eqref{nonkah} 
are ($F_1, F_2, F_3, F_4$) respectively with the constraints \eqref{ghonti}, \eqref{whatif}, \eqref{tnone}, \eqref{lerae} and \eqref{addcons}. 
As mentioned, its equally easy to work out the generic case with the warp factors functions of the angular coordinates ($\theta_1, \theta_2$) in addition to
$r$, but we
will not do so here. In fact this will be left as an exercise for the reader.

\subsection{Towards explicit solutions for the background \label{exsol}}

Let's now consider explicit examples which obey the constraints described in sec. \ref{sec2} (warp factor constraints), sec. \ref{sec3} (torsion constraints), 
and sec. \ref{sec5} (constraint to solve the one-form ODE via separation of variables).  We can organize these constraints in a relatively simple way, to provide a systematic 
method to generate solutions for warp factors and dilaton. 

We start with the special-Hermitian manifolds given as \eqref{kolaram}, with a 
certain appropriate functional form for $F_2(r)$ that is well defined in the regime $0 \le r < \infty$. With this 
input form for $F_2$ we can determine the functional form for $g_1(r)$ satisfying \eqref{lerae} and \eqref{addcons} using the following
differential equation:
\bg\label{godadhor}
{dg_1 \over dr} = {(1+g_1)\left(\phi_r+ {H_r\over 2H}\right) + {2g_1e^{-2\phi}\over F_2}\left(3\phi_r + {H_r\over 2H} + {F_{2r}\over F_2}\right) \over
\left(2 + {3e^{-2\phi}\over F_2}\right)} 
\equiv {2g_1\over F_2 (3+2F_2)}{dF_2\over dr} . \nd
In the second equality we have inserted the background choice \eqref{hdelta} which is appropriate for the special-Hermitian manifolds, where
both the dilaton and $H$ are constants. 

Once we determine $g_1$ from above, we can determine the warp factor $F_1$ using the functional forms for ($g_1, F_2$) via the following equation:
\bg\label{F1eq}
F_1 = \left({dg_1\over dr}\right)^2 \cdot \left({e^{-2\phi}\over F_2 H}\right) \equiv {1\over F_2}\left({dg_1\over dr}\right)^2  , \nd
where the boundary values for the warp factors may be specified by demanding asymptotic regularity along the radial direction. Thus we have a solvable system, requiring only the input functional forms for $F_2$. We will discuss additional constraints coming from quantization of 
${\cal F}_3$ fluxes from the sources on the underlying manifold \eqref{nonkah}. 

We can now solve all the constraint equations with vanishing dilaton i.e $e^{\phi_0} = 1$ and implementing the ODE constraint. A closed form expression can be found for all the 
warp factors in terms of the input function $F_2(r)$. We will first need $g_1(r) = -\vert g_1(r)\vert$. This is given by:
\bg\label{g1r}
\vert g_1(r)\vert = \left[{F_2(r)\over 3 + 2F_2(r)}\right]^{2/3}. \nd
Clearly by construction $\vert g_1\vert < 1$ implying that the resolution parameter $(1 - \vert g_1\vert)F_3$ is always positive definite. 
Let us assume that $F_2(r)$ is a monotonically increasing function of $r$ such that ${\rm min}(F_2) \ll 3/2$. 
In this case $\vert g_1\vert$ is bounded as:
\bg\label{bound}
\left({ {\rm min}(F_{2}) \over 3 }\right)^{2/3} ~\le~ \vert g_1\vert ~ \le ~ \left({1\over 2}\right)^{2/3}.\nd 
Interestingly, if ${\rm min}(F_{2}) \gg 3/2$, we can ignore the 3 in the denominator of \eqref{g1r} for all values of $F_2$, and $\vert g_1\vert$ then takes the following approximate value for
all $r$:
\bg\label{chapdas}
\vert g_1 \vert ~ \approx ~ 2^{-2/3} , \nd
in appropriate units\footnote{We haven't kept track of the units, but they can be inserted in by the careful and diligent readers.}. This means that the resolution parameter 
is approximately $0.37$ times the warp factor $F_3$.

Once we have the functional form for $g_1(r)$ in terms of the input function $F_2(r)$, we can determine the rest of the warp factors satisfying the constraint equations \eqref{godadhor},
\eqref{ghonti}, \eqref{whatif} and \eqref{lerae}. They are now expressed as:
\bg\label{sol1} 
&& F_1 =\frac{4 F_{2r}^2}{F_2^{5/3} (3 + 2F_2)^{10/3}}, ~~~F_3 = 1 -\frac{2+ F_2}{F_2^{1/3} (3 + 2F_2)^{2/3}} , \nonumber\\
&&\;\;\;\; F_4= \left[1 -\frac{2 + F_2}{F_2^{1/3} (3 + 2F_2)^{2/3}}\right]{F^{2/3}_2\over (3 + 2F_2)^{2/3}}, \nd
where again $F_{2r} = dF_2 /dr$. Note that we also require the warp factors to be positive definite, which can act as an additional constraint on the warp factors. Since $F_1$ is positive definite, the requirement that $F_3$ be positive definite is:
\bg\label{conF2}
\left(2 + {3\over F_2}\right)^{2/3} ~ > ~ \left(1 + {2\over F_2}\right). \nd
For a monotonically increasing function $F_2$ with large min$(F_2)$, this inequality is automatically satisfied. For small values of $F_2$,  i.e. at small $r$, this inequality becomes hard 
to satisfy unless:
\bg\label{minua}
{\rm min}(F_2) ~ \geq ~ 2. \nd 
Unfortunately this doesn't quite match-up with the lower bound of $\vert g_1\vert$ discussed in \eqref{bound}, although it is closer to \eqref{chapdas}. Taking \eqref{minua} into account, 
\eqref{bound} changes to:
\bg\label{shalashuor}
\left({2\over 7}\right)^{2/3} ~ \le ~ \vert g_1 \vert ~ \le ~ \left({1\over 2}\right)^{2/3}. \nd
Therefore with \eqref{shalashuor} and \eqref{minua} in mind, a generic monotonically increasing functional form for $F_2$ can be constructed in the following way:
\bg\label{anforF2}
F_2(r) ~ \equiv ~ p(r) + 2 q(r) , \nd
with $p(r)$ and $q(r)$ as two monotonically increasing functions such that $p(0) = 0$ and $q(0) = 1$. They cannot be completely arbitrary as
the quantization of \eqref{dcal} from the 
D5-brane source will relate the coefficients of $p(r)$ and $q(r)$.  

Therefore with the choice \eqref{anforF2} for $F_2(r)$, we can determine all the warp factors \eqref{sol1}. This class of solutions 
corresponds to a class of supersymmetric resolved conifold solutions \eqref{nonkah}, on non-K\"ahler special-Hermitian manifolds. 
After lifting to M-theory, performing a boost, and dimensionally reducing back to IIB, they give a class of IIB solutions describing delocalized five-brane sources 
on resolved conifolds, which are again complex, non-K\"ahler, and supersymmetric. On the other hand if we take more generic $F_2$, not necessarily monotonically increasing, we
can still find solutions with postive definite warp factors. Additionally, if we relax the ODE constraint \eqref{godadhor}, e.g. by resorting to gauge-field choices \eqref{kalabeta} and 
\eqref{kalasika}, more solutions could be found satisfying the constraints \eqref{ghonti}, \eqref{whatif} and \eqref{lerae}. 


With the solution \eqref{sol1} we are ready to determine the one-form.
The one-form $\zeta$ can then be worked out from equation \eqref{haro}, and is given by,
\bg\label{haronow}
\zeta = g_0~ \sin~\theta_2~A_\zeta 
 \left(d\Psi - \vert g_1\vert ~\cos~\theta_2 ~d\phi_2\right), \nd  
where $g_0$ is the normalization constant and the two-form $\omega$ is given by $d \zeta$. 
The functional form for $A_\zeta$ is given as: 
\bg\label{azeta}
A_\zeta = {\rm exp}\left[-\int^r \left({2F_{2r}\over F_2(3+2F_2)}\right) \cdot {dr \over (3+2F_2)^{2/3} F_2^{1/3} - 2 - F_2}\right]. \nd
The asymptotic behavior for $A_\zeta$ can be easily determined using the monotonically increasing function $F_2(r)$. For large values of $F_2$, i.e. at large $r$, 
$A_\zeta$ approaches the following limit:
\bg\label{chokac}
A_\zeta ~ = ~ {\rm exp}\left[\left(2\over 2^{2/3} -1\right)\cdot {1\over F_2^2}\right] ~ \to ~ 1. \nd 
Thus both $\mathcal{A}_{\zeta}$ in \eqref{azeta} and $g_1$  in \eqref{g1r} approach a constant at 
$r \rightarrow \infty$, and hence $\omega \rightarrow 0$ asymptotically, as is required for $\omega$ to be normalizable. Interestingly, at the origin $r \to 0$ again 
$A_\zeta$ and $g_1$ approach constant values and therefore $\omega$ vanishes. 
This is different from the blow-up behavior that we saw for the earlier cases and is perfectly consistent with the fact that there 
are only delocalized sources for this background: thus no singularities from localized branes.

Our next example is a more interesting one because of non-trivial dilaton profile, and is given by the metric \eqref{dhonsho} and three-form fluxes \eqref{altex}. In the language
of \eqref{kunku} the warp factors $F_i$ for \eqref{dhonsho} can be expressed in terms of a function $F(r)$ as:
\bg\label{fachanow}
F_1 = {1\over 2F}, ~~~~~ F_2 = {r^2 F\over 2}, ~~~~~F_3 = {r^2\over 4} + a^2 e^{-2\phi}, ~~~~~ F_4 = {r^2\over 4}. \nd
We have developed the analysis in the previous section, so we will be brief here. To start we will need the gauge field from \eqref{biratpa}. This is given by:
\bg\label{marpa}
{\cal A}_1 = {\rm cosh}~\beta ~ \widetilde{g}_1(r)~\cos~\theta_2~d\phi_2 , \nd
where $\widetilde{g}_1$ can be read off from \eqref{biratpa} as:
\bg\label{pacharrag}
\widetilde{g}_1(r) = - 2 a^2 r F(r)~ {d\phi\over dr}, \nd
which vanishes when either the resolution parameter vanishes or the dilaton is constant. As before quantization of \eqref{biratpa} will 
constrain somewhat the functional form for $F(r)$.
In the limit of vanishing resolution parameter but
non-vanishing dilaton profile, the internal manifold in 
\eqref{dhonsho} is still non-K\"ahler because \eqref{kahrest} is not satisfied and hence $dJ$ in \eqref{dj} is non-zero. However both ${\cal H}_3$ and ${\cal F}_3$ 
\eqref{altex} tend to 
vanish for this background, and hence the non-K\"ahlerity is supported purely by the dilaton profile. Furthermore, putting branes in this background
would break supersymmetry exactly like in Pando Zayas-Tseytlin \cite{pandoz}.

 However, for non-zero resolution parameter, we can support branes because the three-form fluxes in \eqref{altex}
do not vanish. In this case, which is the focus of the present work, we allow the following ansatze for the one-form $\zeta$:
\bg\label{zetpac}
\zeta = \widetilde{g}_2(r, \theta_2) \left(d\Psi + \widetilde{g}_1~\cos~\theta_2 ~d\phi_2\right). \nd
The coefficient $\widetilde{g}_2 (r, \theta_2)$ must satisfy constraint equations similar to \eqref{condg2}, and therefore we allow for separation of variables as in \eqref{ghatot}. 
We would face a similar subtlety as in \eqref{subtle}, unless we allow:
\bg\label{addconow}
{\partial \widetilde{g}_1\over \partial r} = - 2 a^2 \sqrt{G_3\over G_6}\nd
which differs from \eqref{addcons} by the coefficient $a^2$. This coefficient is essential because to zeroth order in $a^2$, $\widetilde{g}_1$ in \eqref{pacharrag} vanishes, 
whereas ($G_3, G_6$) do not. This is consistent with the fact that to zeroth order in $a^2$ we do not expect a Taub-NUT space in M-theory for the supersymmetric case. 

On the other hand, the radial part in the decomposition \eqref{ghatot} must satisfy an equation similar to \eqref{lattobaaz}. 
The condition \eqref{addconow} leads to the following constraint on the warp factors:
\bg\label{lengtam}
{d\over dr}\left(r F {d\phi\over dr}\right) = {1\over 2} re^\phi \sqrt{e^{2\phi} {\rm cosh}~\beta - {\rm sinh}~\beta} , \nd
which is in addition to the constraint equation \eqref{thchini} on the warp factors from the supersymmetry conditions. Thus after the dust settles, the one-form $\zeta$ will 
now be given by:
\bg\label{meyer}
\zeta = \widetilde{g}_0~ a^2~ \sin~\theta_2~{\rm exp}\left[-\int^r ~{4a^2\over r}\cdot{\phi_r\over \left(rF\phi_r\right)_r} dr\right]\left(d\Psi 
+ \widetilde{g}_1 \cos~\theta_2~d\phi_2\right) , \nd
where $\widetilde{g}_0$ is a constant independent of the resolution parameter $a^2$, and the subscript $r$ denote derivative with respect to the radial coordinate $r$. 

Note that
if we chose the gauge field ${\cal A}_1$ such that it satisfies:
\bg\label{goromc}
{\cal F}_2 = d{\cal A}_1 + \left(1 + \widetilde{g}_1\right) {\rm cosh}~\beta~e_{\theta_2} \wedge e_{\phi_2} , \nd
instead of \eqref{marpa}, then the only constraint on the warp factor would be \eqref{thchini}, i.e. \eqref{lengtam} will not apply, similar to what we saw earlier. In the absence of
\eqref{lengtam} we will require input functions for the dilaton $e^\phi$ or the warp-factor $F(r)$. 

Finally, we note that in the case of a vanishing resolution parameter (i.e. a singular conifold), a wrapped D5-brane solution \emph{could} be constructed with ISD fluxes, but the resulting construction may not be supersymmetric. This scenario can be easily rectified by altering slightly the warp factor choices in \eqref{fachanow} in the following way:
\bg\label{chinpacha}
F_1 = {e^{-\phi}\over 2F}, ~~~~~ F_2 = {e^{-\phi}r^2 F\over 2}, ~~~~~F_3 = { e^{-\phi}r^2\over 4} + {\cal O}(a^2), ~~~~~ F_4 = {e^{-\phi}r^2\over 4}. \nd
The disadvantage of this approach is that, in the zeroth order in $a^2$, we will have components of three-form fluxes along both ($\theta_1, \phi_1$) and ($\theta_2, \phi_2$) 
directions. The results \eqref{marpa} and \eqref{pacharrag} will appear in the next order in $a^2$. An analysis of this case can be easily performed along the lines of the previous
section, but we will not do it here.

\section{Discussion of Localized Fluxes \label{disc}}

We have considered two classes of examples. In the first class, we have a D6-brane with and without without background fluxes. In the second class, we have a D5-brane wrapped on the two-cycle of a non-K\"ahler resolved conifold in the presence of fluxes, with and without a dilaton profile.  
In the latter category of the second class, i.e the one without a dilaton profile, the sources only appear as delocalized. In the former category of the second class, i.e the one with a dilaton profile, many non-trivial
localized supersymmetric solutions can be constructed.

In each of the two classes, the brane world-volume theory is encoded in the M-theory geometry via the normalizable two-form $\omega$. 
However, we have not considered the effect of non-zero DBI gauge fields on the brane, which would arise in M-theory via the term:
\begin{equation}
{\cal G}_4 = {\cal F} \wedge \omega ,
\end{equation}
where ${\cal F}=d{\cal A}$ is the U(1) field strength of the gauge theory localized on the brane. Since $\omega$ quickly decays to zero
 away from the brane, we refer to fluxes of the above form as {\it localized fluxes}. These have been mentioned in many previous works, see for example \cite{imamura, ashoke, Sen:1994yi, robbins}, but no detailed discussion has yet been presented. These fluxes are intimately related to de Sitter solutions in string theory, as they are the key ingredient in D-term uplifting  
\cite{renata}, which makes use of the D-term potential later derived by Louis and Jockers in  \cite{0502059}, and Haack et al. in \cite{Haack:2006cy}.
 
Up to this point all of our examples have implicitly assumed zero localized flux, i.e. ${\cal F}=0$. 
In particular, we have only studied our solutions at lowest order in $\alpha'$, such that the bulk IIA RR field ${\cal F}_2 = d {\cal A}_1$ does not induce any 
non-zero ${\cal F}=d{\cal A}$ on the brane. We would like to study the effect of including non-zero localized fluxes, as one would expect ${\cal F}$ 
to be non-zero in a generic flux compactification ignoring, for the moment, the backreaction of the localized flux. That is, we will keep the background solution (specified by the metric and non-localized flux) fixed, and consider the effect of including a small localized flux ${\cal F} \wedge \omega$.  
In an upcoming work, we will consider the full solution, including the corrections to the metric and complex structure.
 
Our strategy to discuss the localized flux in M-theory relied on the existence of 
a one-form $\zeta$, from which the harmonic two-form $\omega=d\zeta$. For the first 
class of examples, we see that $\zeta$ takes the following form, up to possible scalings:
\bg\label{cat1}
\zeta = g(r) \left(d\Psi + \cos~\theta_2~d\phi_2\right)\nd
where $g(r)$ takes the form \eqref{zetbeta} for the vanilla (i.e. flux less) D6-brane and \eqref{churi} for the case of D6-brane with fluxes. 

For the second class of examples, the one-form $\zeta$ is more non-trivial and takes the following form:
\bg\label{japnunu}
\zeta = g_2(r, \theta_2)\left[d\Psi + g_1(r)~\cos~\theta_2~d\phi_2\right]\nd 
where $g_1(r)$ is a non-trivial function of the radial coordinate and is given by \eqref{g1r} for the constant dilaton case and by \eqref{pacharrag} for the case with a 
dilaton profile. The other function $g_2(r, \theta_2)$ takes the generic form \eqref{haro}, and is given by \eqref{haronow} for the constant dilaton case, and by \eqref{meyer} for
the case with a dilaton profile. 

At this stage, we can study the M-theory picture either in terms of a four-fold ${\cal M}_8$
with a $SU(4)$ structure or in terms of a seven-dimensional manifold ${\cal M}_7$ with a $G_2$ structure. In the
former case the four-fold will be parametrized by coordinates: ($\theta_1, \phi_1$), ($\theta_2, \phi_2$), ($r, \psi$), ($x_{11}, x_3$), whereas in the latter case the 
seven-dimensional manifold will be parametrized by coordinates 
($\theta_1, \phi_1$), ($\theta_2, \phi_2$), ($r, \psi, x_{11}$). The seven-dimensional manifold with $G_2$ structure is locally a Taub-NUT space oriented along 
($\theta_2, \phi_2, r, x_{11}$) fibered over a three-dimensional base parametrized by ($\theta_1, \phi_1, \psi$). In the language of the ${\cal M}_7$, the G-flux \eqref{lrus}
can be re-written as:
\bg\label{lrusnow}
{{\cal G}_4\over {\rm sinh}~\beta} &=& \left(\sqrt{F_1 F_2} - F_{3r}\right) e_r \wedge e_{\theta_1} \wedge e_{\phi_1} \wedge e_{11} + \widetilde{e}_\psi \wedge e_{\theta_1} \wedge e_{\phi_1}\wedge \widetilde{e}_{11} \nonumber\\
&+& \left(\sqrt{F_1 F_2} - F_{4r}\right) e_r \wedge e_{\theta_2} \wedge e_{\phi_2} \wedge e_{11} + \widetilde{e}_\psi \wedge e_{\theta_2} \wedge e_{\phi_2}\wedge \widetilde{e}_{11}\nd
where the new vielbeins are defined in the following way:
\bg\label{viludef}
\widetilde{e}_\psi = d\psi, ~~~~~ e_{11}= dx_{11} + {\cal A}_1, ~~~~~ \widetilde{e}_{11} = dx_{11} + {\cal A}_3\nd
with ${\cal A}_1$ and ${\cal A}_3$ are given in \eqref{cala1} and \eqref{kalasika} respectively. Note that the way we constructed the fluxes, they are naturally 
defined on ${\cal M}_7$ instead of ${\cal M}_8$. The flux \eqref{lrusnow} does not have a $x_3$ component, so if we use a four-fold, we cannot define the self-duality 
naturally, although \eqref{lrusnow} is supersymmetric by construction.


The localized flux for the vanilla D6-brane, on the other hand, can easily be made self-dual when defined on a four-fold, as is 
required to satisfy the equations of motion for a compactification to (warped) Minkowski space, 
see for example \cite{DRS, GKP}.  This occurs because the gauge field ${\cal F}$ is necessarily transverse to the Taub-NUT space, 
while $\omega = d\zeta$ lives strictly \emph{on} the Taub-NUT space. Hence the dual of ${\cal G}_4$ on the fourfold is given by:
\begin{equation}
\ast_{\scalebox{.45}{8}}~ {\cal G}_4 =  \ast_{\scalebox{.45}{\rm NT}}~ {\cal F} \wedge \ast_{\scalebox{.45}{\rm TN}}~ \omega,
\end{equation}
where we have denoted the hodge star on the Taub-NUT directions with a subscript TN, and the remaining four orthogonal directions as NT. 
The above equation is satisfied for self-dual ${\cal F}$, and thus the equations of motion can be satisfied without resorting to additional effects such 
as the generation of a cosmological constant. One may also check that this flux will not break supersymmetry, 
namely that it is primitive with respect to the simplest choice of complex structure on the manifold with metric given by equation (\ref{reg7}). 

The above argument extends straightforwardly to the D6-brane in a flux background, although the Taub-NUT space becomes slightly deformed due to the non-zero twist parameter $\alpha$ encoding
 the presence of fluxes. The M-theory fourfold is split into a Taub-NUT and four orthogonal directions, and the normalizable two-form that spans the Taub-NUT 
space can be computed by demanding self-duality and normalizability. As before, the introduction of localized flux will not break supersymmetry and will not generate 
a cosmological constant, provided suitable conditions are placed on ${\cal F}$.

The second case that we study here, however, cannot be lifted in M-theory on a 
four-fold because any duality to M-theory will convert the wrapped D5-brane into an M-theory five-brane and not into geometry. Thus the seven-dimensional manifold with $G_2$
structure, which in-turn is a 
warped Taub-NUT space fibered over a three-dimensional base, is probably the best way to analyze this picture. However, if we allow an additional D7-brane in the type IIB side, a
four-fold description will become useful. In fact this is also where we can study D-term uplifting \cite{renata}.

\section{Conclusion and Future Directions \label{concl}}

In this work we have constructed explicit supersymmetric solutions for D5 branes wrapping a resolved conifold. We accomplished this by duality chasing a supersymmetric conifold solution with general warp factors, and only $\mathcal{H}_3$ flux, to a solution with $\mathcal{H}_3$, $\mathcal{F}_3$, and $\mathcal{F}_5$ flux, as well as five-brane sources. In this way, supersymmetry is built in to the final solution, provided we satisfy certain constraints on the warp factors, which we have verified explicitly. 

Interestingly, both the ``before duality'' and ``after duality'' solutions are non-K\"ahler, but the detailed properties depend intimately on the dilaton: a constant dilaton corresponds to a special-Hermitian manifold which dualizes to a complex manifold with delocalized sources, while a non-constant dilaton $\phi=\phi(r)$ corresponds to a complex manifold which dualizes to a non-complex manifold with localized sources (i.e. D5 branes). The solutions we found have the following form for the metric, equation \eqref{iibform},
\bg
&&ds^2 = {1\over e^{2\phi/3} \sqrt{e^{2\phi/3} + \Delta}} ~ds^2_{0123} + e^{2\phi/3} \sqrt{e^{2\phi/3} + \Delta} ~ds^2_6 ,  \\
&& ds^2_6 = F_1~ dr^2 + F_2 (d\psi + {\rm cos}~\theta_1 d\phi_1 + {\rm cos}~\theta_2 d\phi_2)^2  + \sum_{i = 1}^2 F_{2+i}
(d\theta_i^2 + {\rm sin}^2\theta_i d\phi_i^2) . \nonumber 
\nd
We derived a class of solutions with constant dilaton, with warp factors given by:
\bg
&& F_2 = F_2(r), \;\;\;\;\;\; \phi=\phi_0 , \\
&& F_1 =\frac{4 F_{2r}^2}{F_2^{5/3} (3 + 2F_2)^{10/3}}, ~~~F_3 = 1 -\frac{2+ F_2}{F_2^{1/3} (3 + 2F_2)^{2/3}} , \nonumber\\
&&F_4= \left[1 -\frac{2 + F_2}{F_2^{1/3} (3 + 2F_2)^{2/3}}\right]{F^{2/3}_2\over (3 + 2F_2)^{2/3}} ,\nonumber \nd
where $F_{2r}= dF_2/dr$, in terms of a general $F_2(r)$, as well as a class of solutions with varying dilaton, given by
\bg\label{jolontop}
F_1 = {1\over 2F}, ~~~~~ F_2 = {r^2 F\over 2}, ~~~~~F_3 = {r^2\over 4} + a^2 e^{-2\phi}, ~~~~~ F_4 = {r^2\over 4}, ~~~~~ \phi=\phi(r) , \nd
for a general function $F(r)$. In \eqref{fachanow} we also briefly discuss a slight variant of \eqref{jolontop} given by the dilaton factor.

We then studied the M-theory lift of these solutions, and related examples. The worldvolume theory of the brane is encoded into a normalizable two-form $\omega=d\zeta$ which describes the M-theory geometry around the position of the brane. In simple cases, such as a D6 brane, the M-theory geometry is a simple Taub-Nut space embedded in a fourfold, and the calculation of $\omega$ is straightforward. For the solutions described above, the lift to a M-theory is much less trivial; while we were able to compute the normalizable two-form, a meaningful discussion of localized fluxes (which encode the DBI gauge field on the brane) requires studying M-theory on a seven-dimensional manifold with $G_2$ structure. We leave this discussion, as well as an extension to include D7 branes, to future work.

Despite this apparent set-back, we have achieved quite a bit using our M-theory construction. The extra constraints on the warp factors of 
non-K\"ahler resolved conifold
\eqref{nonkah}, that appear from analyzing 
M-theory harmonic forms are useful for predicting certain interesting geometric behaviors of these manifolds. There are a large class of these manifolds admitting supersymmetric fluxes
that are useful for studying new aspects of string compactifications. We have simply scratched the surface. 


\section*{Acknowledgements}

We would like to thank Stanford Physics department for providing a stimulating environment and hospitality during the course of this work. The work of K. D is supported in part by
the National Science and Engineering Research Council of Canada and in part by the Simons Research Grant. The work of E. M is supported by the National Science and Engineering Research Council of Canada via a PGS D fellowship.

{}
\end{document}